\documentclass[12pt,preprint]{aastex}
\usepackage{graphicx}
\usepackage{ulem}
\usepackage{natbib}

\usepackage{color}

\definecolor{red}{rgb}{0.7,0,0}
\definecolor{blue}{rgb}{0,0,0.7}

\title{A Temporal Analysis Indicates a Mildly Relativistic Compact Jet in GRS~1915+105}
\author{Brian Punsly\altaffilmark{1}} \and\author{J\'{e}r$\hat{\mathrm{o}}$me Rodriguez\altaffilmark{2}}
\altaffiltext{1}{1415 Granvia Altamira, Palos Verdes Estates CA, USA
90274 and ICRANet, Piazza della Repubblica 10 Pescara 65100, Italy,
brian.punsly1@verizon.net} \altaffiltext{2}{Laboratoire AIM,
CEA/DRF-CNRS-Universit\'{e} Paris Diderot, IRFU SAp, F-91191
Gif-sur-Yvette, France.}
\begin{document}
\begin{abstract} Most of our knowledge of the radio morphology and
kinematics of X-ray binary partially synchrotron self-absorbed
compact jets (hereafter compact jets) is based on the observations
of GRS~1915+105 which has the most prominent compact jet. Yet, the
compact jet bulk velocity, $v$, is poorly constrained in the
literature, $0.07 < v/c < 0.98$. In spite of this uncertainty,
compact jets are often unified with relativistic jets in active
galactic nuclei. We have estimated $v$ as part of a temporal
analysis of GRS~1915+105 jets in ``high plateau states" (HPS). We
define the HPS as a state showing a hard X-ray spectrum and low
level of long-term ($>10$s) X-ray activity associated with 15 GHz
flux density $>70$ mJy for $>7$ consecutive days. The radio emission
is associated with compact jet emission. Two HPS were monitored at
15 GHz during their termination with e-folding times of 3.8 hrs and
8.6 hrs. We combine this time scale with the scale of spatial
variation of the linear source of a VLBA image preceding the fade of
one of these HPS in order to estimate the jet speed. Our assumption
that the reduction in radio emissivity propagates as an approximate
discontinuity down the HPS jet (leaving a weak jet in its wake)
indicates $0.17 <v/c< 0.43$. This agrees closely with the only other
existing $v$ estimates that are derived directly from radio images,
jet asymmetry produced by Doppler enhancement.
\end{abstract}
\keywords{Black hole physics --- X-rays: binaries --- accretion, accretion disks}

\section{Introduction}
Galactic black hole X-ray binaries (BHXRB) appear as compact flat
spectrum radio sources during X-ray hard states
(\citet{cor00,cor03,gal03}), that is widely interpreted as partially
synchrotron self-absorbed compact jets (hereafter compact jets; e.g.
\citet{kai06}). The ``compact jet" is suppressed during transition
to the soft X-ray states \citep{fen98}. GRS 1915+105 has the most
prominent compact jet, an order of magnitude more luminous than that
of other BHXRB.  Thus, it is the only compact jet with a rich
history of VLBI observations \citep{dha00,fuc03,rib04}. As opposed
to conical SSA jets, the properties of GRS~1915+105 are better
explained with a structured jet (see e.g. \citet{pun10}). The X-ray
luminosity of GRS 1915+105 is also one of the highest of any known
Galactic black hole candidate \citep{don04}. The existence of both a
jet and high continuum luminosity make it tempting to speculate that
GRS 1915+105 might be a scaled down version of a radio-loud quasar.
This speculation has been further fueled by the highly publicized
relativistic discrete ejections \citep{mir94,fen99}.

The compact jet is one of three distinct radio phenomenon associated
with GRS~1915+105. In this article, we emphasize the distinct
character of these radio states. Much clarity has been attained in
the classification of radio states by considering the X-ray evidence
associated with each type of radio emission.
\begin{enumerate}
\item Major superluminal, discrete ejections are preceded by a hard X-ray
flare that begins a few hours before jet launching. The strength of
the X-ray flare is correlated with the kinetic luminosity of the
ejection and are a sizeable fraction of the Eddington luminosity for
powerful flares. During the jet launching, the X-ray flux is highly
variable with a time average value roughly equal to the pre-flare
level \citep{pun11}.
\item The continuous compact jet is accompanied by a hard X-ray
state that is steady. Using the classifications of \citet{bel00}, a
$\chi$ class of X-ray variability state coexisits with the compact
jet \citep{kle02}
\item Quasi-periodic radio states are associated with oscillations of the X-ray light curves that are near their nadir when the ejection
occurs. These are highly variable X-ray luminosity classes, showing
fast transitions between hard and soft, very intense states. These
states oscillate between low/hard and high/soft with the transition
between low/hard and high/soft marked by an X-ray spike, that is the
moment of the discrete radio ejection. These radio oscillations
typically have peaks of 10 mJy - 50 mJy in the radio
\citep{kle02,rod09,poo97}. It was shown in \citet{rod09} that a
bright soft spike preceded each ejection similar to the discrete
ejections described in (1). It should be noted that these
quasi-periodic events include weak optically thin bubbles of radio
emission (a few mJy), \citet{rod09}, and more powerful optically
thick oscillations \citep{mir98,dha00}.
\end{enumerate}

The luminous compact jets, being more ubiquitous than luminous
relativistic discrete ejections in GRS~1915+105, have received most
of the observational attention. Yet, the compact jet has no direct
measurement of jet bulk velocity. This has led to some indirect
estimates of bulk velocity. The primary motivation of our temporal
analysis is to provide an independent determination of the compact
jet bulk velocity derived directly from observation. As a corollary
to this effort, we attempt to bring clarity to the notion of the
continuous steady compact jet. A major source of confusion occurs in
the interpretation of the compact jet as a consequence of radio
imagery. Due to the insufficient u-v coverage and sensitivity of
VLBA (Very Large Baseline Array) observations, the jet during highly
variable (quasi-periodic) modest radio luminosity states ($\sim 15 -
50 $ mJy at 15 GHz) appears to have the same morphology as the jet
in strong ``steady" states ($\sim 100 - 150 $ mJy at 15 GHz), a twin
exhaust flame \citep{dha00}. Thus, the compact nature of the jets in
steady states and during quasi-periodic radio states are often
considered without distinction \citep{dha00}. It is tempting to
combine superluminal discrete ejections, the steady continuous
compact jet and the stronger quasi-periodic radio states in a
unified scheme based on the observational evidence for a common jet
direction in all three cases \citep{dha00}. However, the degree to
which these phenomena can be consolidated into one jet model is a
subject of conjecture and is yet to be established by direct
observational measurement \citep{fen04}. The compact jet is
consistent with continuous radio emission on time scales as short as
tens of seconds and for this reason it is distinct from the
phenomenon that creates the quasi-periodic radio oscillations (see
Section 4). The quasi-periodic radio emission is often assumed to
arise from discrete ejecta, but emission cycles might also be a
consequence of shocks or dissipative MHD wave trains within a jet
\citep{dha04}. The desire to combine the three radio phenomenon in a
unified model has led to speculation about the compact jet bulk
velocity. It has been argued that since superluminal ejecta often
appear before or after a compact jet that the compact jet must also
be relativistic \citep{dha00}. But, this has never been directly
observed and is still a conjecture \citep{rib04}. The case was
further bolstered by the idea that quasi-periodic ejecta would have
to expand at near the speed of light if adiabatic cooling is the
source of the fading of the ejecta \citep{dha00}. However, this
argument is not direct because (as we show in Section 4) the
continuous compact jet shows no quasi-periodic behavior in the radio
\citep{kle02}. Thusly motivated, the main goal of this paper is to
add an independent method of determining the compact jet bulk
velocity directly from observation to our knowledge base.

\begin{figure}
\begin{center}
\includegraphics[width=150 mm, angle= 0]{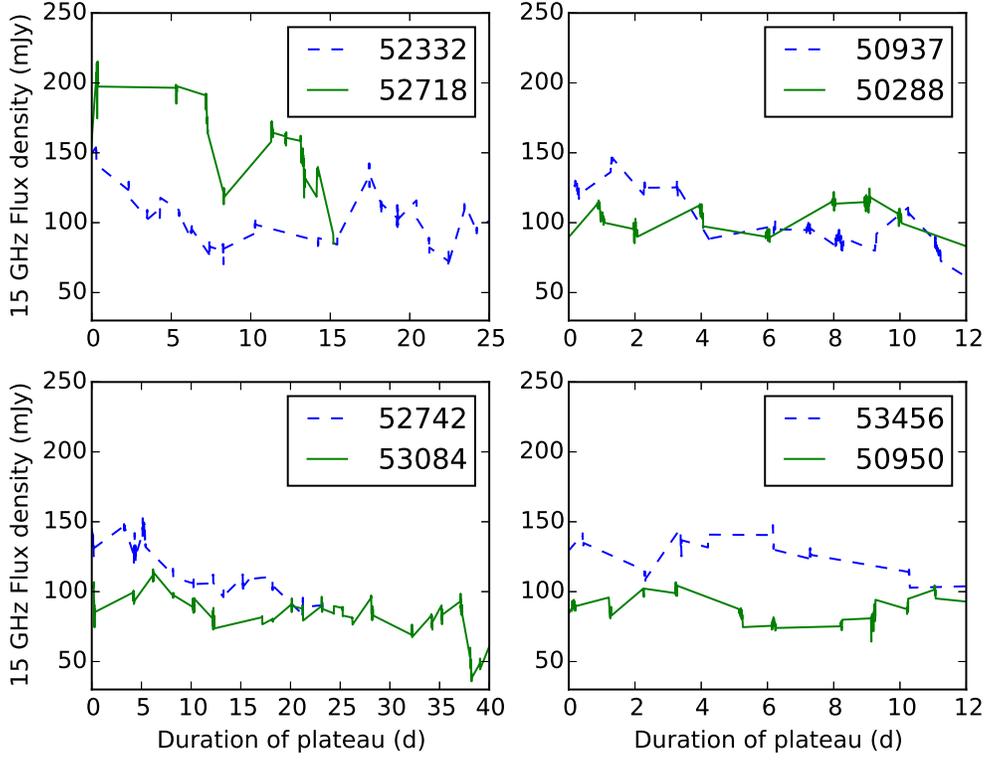}
\caption{The 15 GHz light curves of the 8 high plateau states
studied in this article. These states are very stable for
GRS~1915+105, having modest to small variations on time scales of
hours to a day.}
\end{center}
\end{figure}
\par

\begin{table}
\caption{High Plateau States} {\footnotesize\begin{tabular}{ccccc}
\hline
Start Date (MJD) &  Duration (Days) & Maximum 15 GHz Flux Density\tablenotemark{a} & Minimum 15 GHz Flux Density\tablenotemark{a} \\

\hline
50288   & 9  & 118 mJy & 93 mJy   \\
50937   & 11  & 121 mJy & 87 mJy   \\
50950  & 12  & 105 mJy  & 79 mJy  \\
52332  & 28  & 139 mJy   & 73 mJy   \\
52718 & 15  & 209 mJy & 75 mJy    \\
52742  & 25  &  144 mJy  & 89 mJy   \\
53084   & 37  & 111 mJy & 71 mJy  \\
53456   & 10  & 141 mJy & 109 mJy   \\

\end{tabular}}
\tablenotetext{a}{Flux density averaged over duration of the Ryle
telescope observation on that day}
\end{table}

\begin{figure}
\begin{center}
\includegraphics[width=125 mm, angle= 0]{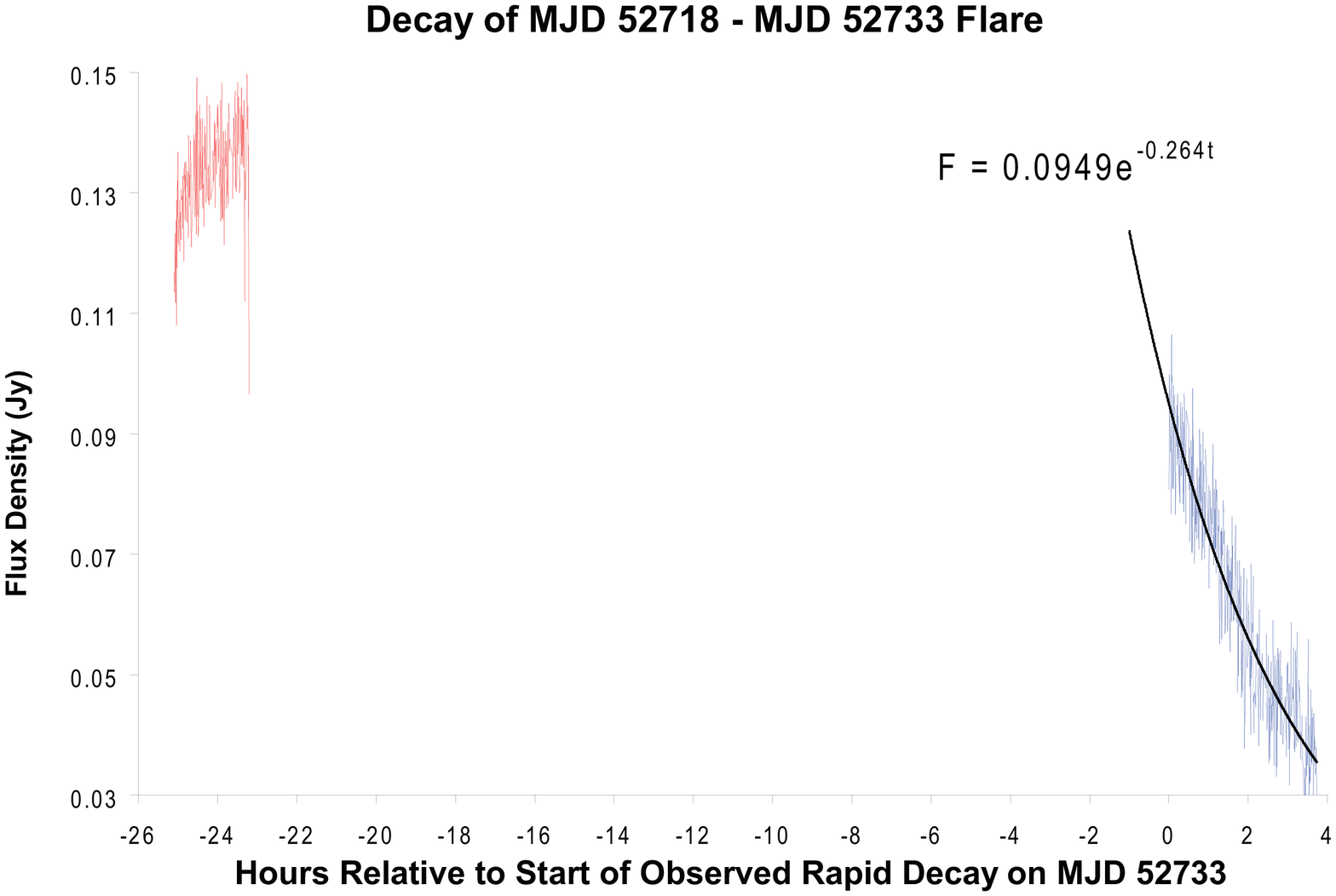}
\includegraphics[width=125 mm, angle= 0]{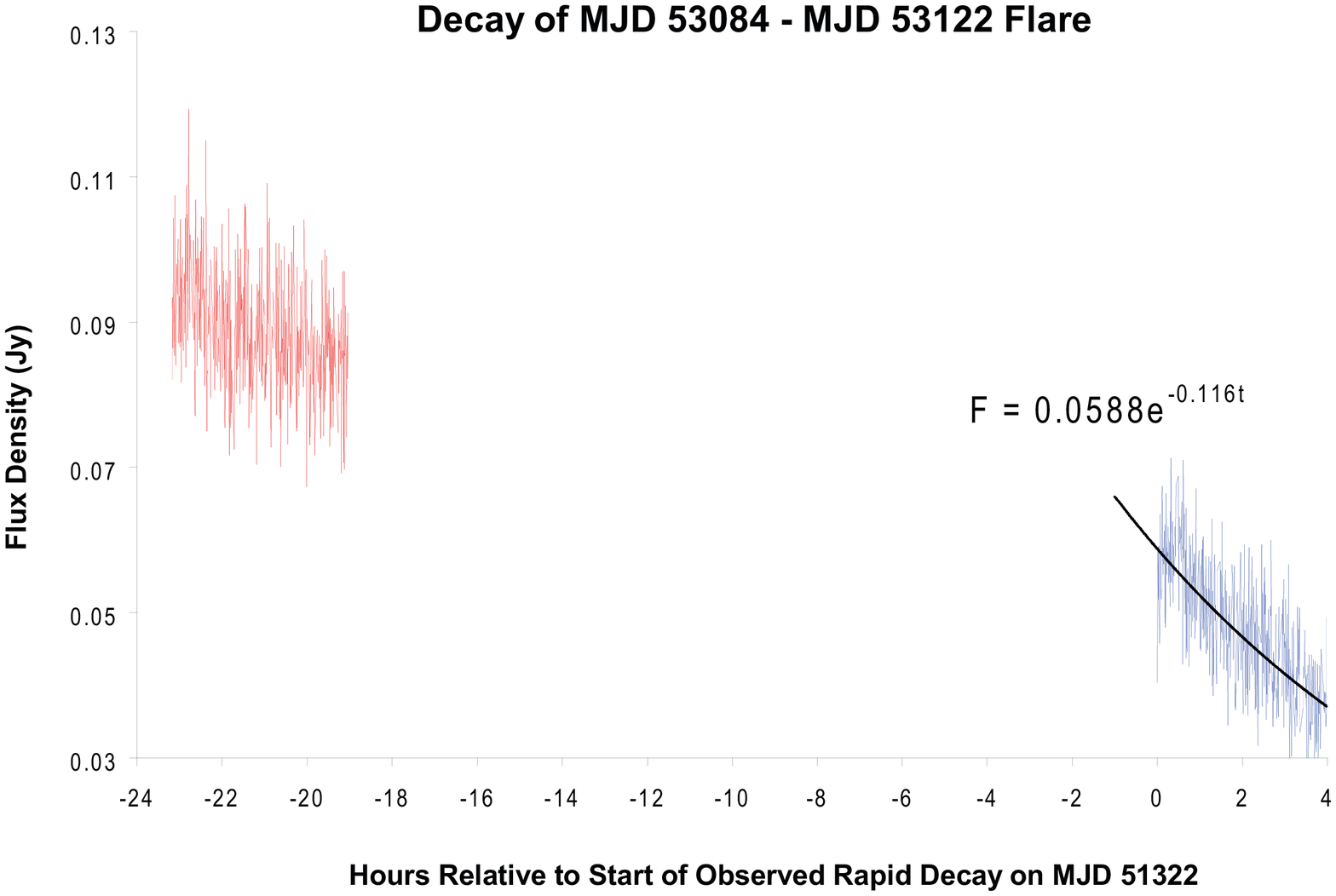}
\caption{The Ryle Telescope was used to produce light curves of the
decay of two of the HPS. The 15 GHz flux density, $F$, is fit by an
exponential function during the decay. The unit of time in the
exponential fit is hours. The decay on MJD 52733 in the top frame
and on MJD 53122 in the bottom frame have e-folding times of 3.8 hrs
and 8.6 hrs, respectively.}
\end{center}
\end{figure}

\par  In this study, we search for some evidence in the radio data that could
provide an indication of the bulk velocity of the compact jet in
GRS~1915+105. Our method considers ``high plateau states" (HPS,
hereafter) that are defined by more than 70 mJy of flux density at
15 GHz for more than 7 days of monitoring. These radio states are
always accompanied by an X-ray $\chi$ state. The high plateau states
are by all accounts strong compact jets and they provide an
excellent background on which to study temporal variations of a
stable structure. The large flux density allows short 32 s sampling
with reasonably good statistics. There are 8 such high plateau
states in the Ryle Telescope 15 GHz archive \citep{rus10}. These 8
states are defined in Table 1. The light curves are shown in Figure
1.
\par In section 2, the bulk compact jet velocity is estimated. By comparing
the time constant, $T$, of the exponential decay of the 15 GHz light
curve that terminates two HPS with the spatial exponential decay
constant, $D$, for the linear source of the VLBA imaged compact jets
in HPS, we estimate the jet velocity, $V=D/T$. In Section 3, we
compare and contrast the X-ray properties during the HPS and during
the exponential decay of the radio light curve. A time series
analysis of the radio data during HPS is presented in Section 4.
These findings are summarized and consolidated into the unified
scheme in the Discussion.
\begin{figure}
\begin{center}
\includegraphics[width=80 mm, angle= 0]{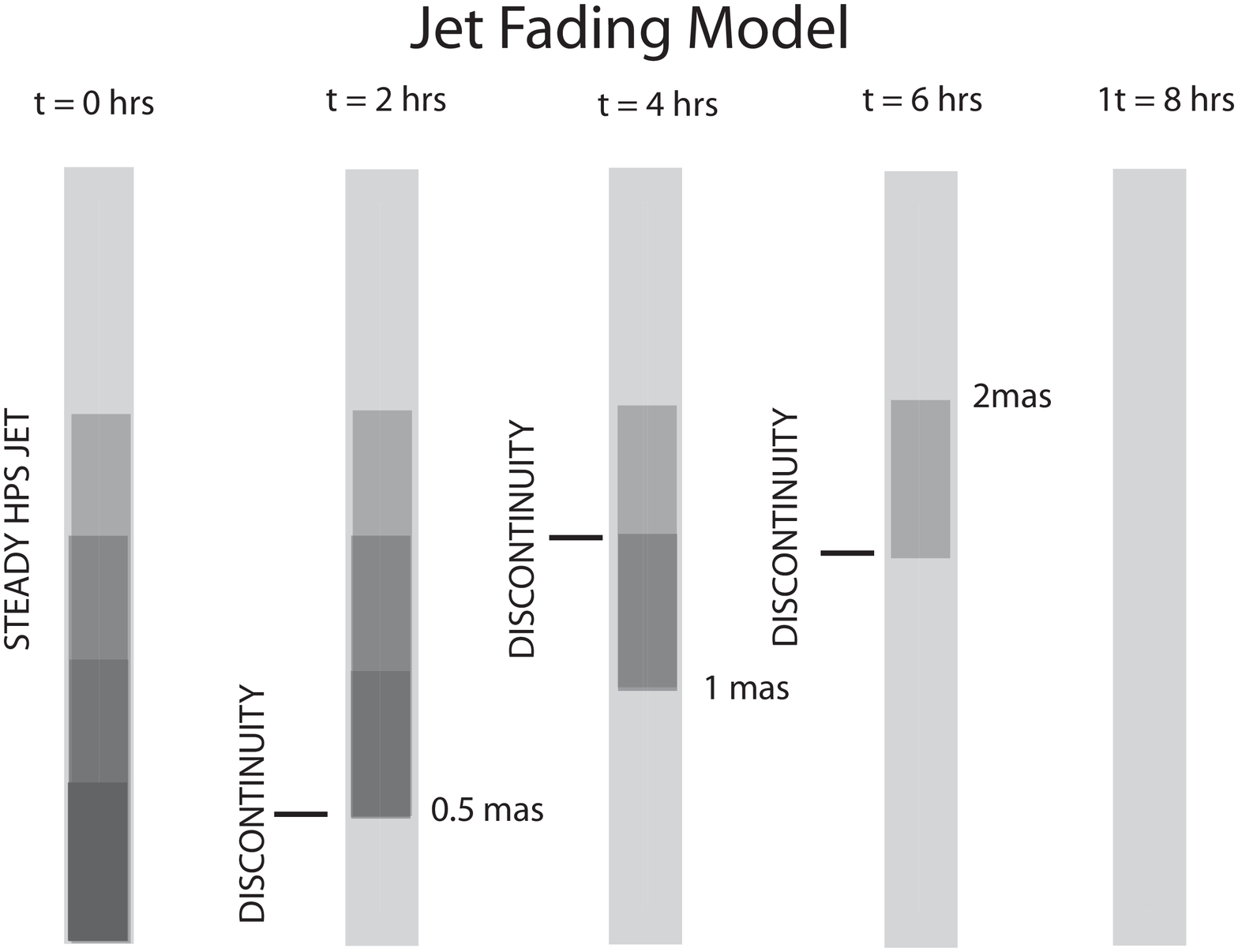}
\caption{The time evolution scenario for jet fading that is
described in Section 2.1. The figure shows a large transient in the
form of a discontinuity that propagates down a conduit formed by the
pre-existing steady state HPS jet. The scales are based on the
analysis of the VLBA maps of jets in HPS that are described in the
text and the light curves in Figure 2. The details of the time
evolution are described in the text Note that there are no VLBA
images of this transient stage and it is a speculative model.}
\end{center}
\end{figure}

\section{Bulk Velocity Estimate}

In this section, we estimate the jet bulk velocity. Our method
combines the timescale for the exponential decay of radio emission
at the end of the HPS state with the characteristic spatial extent
of a HPS jet, as determined in Sect. 2.3. Figure 1 shows the long
term 15 GHz light curves of the eight HPS that have been monitored
with the Ryle telescope \citep{rus10}. There are gaps in the daily
coverage in Figure 1, so the length of each HPS in Table 1 might be
slightly under estimated in some cases. On days when the data was
taken, the source was sampled continuously by the Ryle telescope for
30 minutes to 8 hours. Note that the strong HPS that begins on MJD
52718 is terminated by a decay then a major ejection before another
HPS starts on MJD 52742 and the two events are formally two separate
HPS \citep{fuc04}.

\par Figure 2 shows the serendipitous capture of the decay of two
of the eight HPS in Figure 1. The light curves are well fit by an an
exponential decay. The decay on MJD 52733 in the top frame and on
MJD 53122 in the bottom frame have e-folding times of 3.8 hrs and
8.6  hrs, respectively.

\par In order to determine a bulk flow speed
from these decay times, one needs a measure of the propagation
distance along the jet relative to the core. It is proposed that a
relevant distance can be attained by an analysis of the VLBA images
of jet at the same frequency as the light curves in Figure 2, 15
GHz. The decline in flux indicates that the engine feeding the jet
has either turned off or has been reduced significantly in strength.
After this occurs, one expects that a discontinuity propagates along
the jet leaving a weak jet in its wake. The previously ejected
synchrotron emitting plasma continues to move along the conduit
formed by the steady state HPS jet. There is no apriori reason for
the geometry of the outflow of the previously ejected synchrotron
emitting plasma to be altered from its steady state HPS
configuration to zeroth order. Thus, one should be able to combine
the e-folding distance for spatial decay of the empirically
determined source of the VLBA image, $D$, with the e-folding decay
times, $T$ in Figure 2 to yield an apparent bulk flow jet velocity,
$V_{app}=D/T$. This can be transformed to a physical bulk velocity,
$v$, by the transformation

\begin{equation}
V_{app}= \frac{(v/c) \sin{\theta}}{1-(v/c)\cos{\theta}}\;,
\end{equation}
where $\theta=60^{\circ}$ is estimated from proper motion
measurements of approaching and receding ejecta during an outburst,
combined with the newly-determined parallax distance of 8.6 kpc
\citep{rei14}. The method is described in Section 2.1 and
pictorially by the schematic in Figure 3. The details of an actual
calculation are described in Section 2.3.

\subsection{A Model of Jet Decay}
In this subsection, we motivate a model of the fading compact jet
associated with the decay curves in Figure 2. Note that the fading
compact jet has never been imaged by VLBA. The full force compact
jet has been imaged (see below), but the decay afterwards is
determined only by the light curves in Figure 2. Thus, we need some
information to guide a necessarily speculative model of how a
compact jet fades.

\par In the context of the decay curves in Figure 2, the flux
density decays to low levels, but not zero. Thus, we expect a low
surface brightness jet is the end state. There are some compact jet
states that end with a major ejection. This is in distinction to the
fade seen here. In this case, the compact jet either disappears or
is too weak to be detected with a low sensitivity VLBA observation
\citep{dha04}.

\par The fading of the compact jet or the disappearance of the compact jet is
assumed to represent a decrease in the intrinsic power of the
compact jet. The interior of the jet is hidden by synchrotron
self-absorption (SSA). The emission that is detected at 15 GHz is
the $\tau \sim 1$ surface of the outgoing plasma. We assume that the
power is drastically reduced on a time scale, $T_{0}$, much shorter
than the time it takes the plasma to flow a scale length for the jet
that represents a large fraction of the detected emission with VLBA,
$L$. The abrupt transition to a state of highly reduced jet power
begins at $t = 0$ and proceeds to completion in the time interval $0
< t < T_{0}$. In the examples to follow, $L
> 2$ mas. Namely, if the plasma velocity is $V$, then $VT_{0} \ll
L$. This seems like a reasonable assumption considering that the jet
originates from a region near the black hole that is seven orders of
magnitude smaller than the jet length, $L$ \citep{pun10}. After the
transition, a very weak jet is ejected from the central engine at
$t> T_{0}$. However, there is still plasma from $-L/V< t <0$ that
occupies the conduit of the original HPS jet. The plasma that flows
down this conduit is a strong transient feature that is
approximately a step wave. In the language of magnetohydrodynamics
this is approximately a contact discontinuity (not a shock wave)
that propagates at the speed, $V$, of the bulk flow of the jet
plasma if the plasma is turbulent. The discontinuity might have to
be supplemented by a more complicated rarefaction wave structure in
principle, but this detail will not affect the emissivity and
velocity of the preponderance of the remnant plasma from the HPS
steady jet which is far upstream of the disturbance. The abrupt
change that is assumed to occur is the fundamental assumption of the
model of the jet fading that is manifest as a large transient
discontinuity.
\par The plasma upstream of the propagating discontinuity has virtually the same
plasma properties as the plasma ejected at $t< -L/V$ that was
located at this same axial coordinate ,$x$, along the jet during the
steady HPS jet state. This includes the number density of particles
and the magnetic field strength. The SSA opacity is determined by
the local plasma state and is therefore virtually the same as it was
at $t< -L/V$ during the steady jet phase of the HPS \citep{gin69}.
The region for which this is true changes with time and is defined
by the axial coordinate range, $x
>Vt, \, t >0$. For this portion of the jet, the $\tau \sim 1$ surface
and the intrinsic emissivity is virtually unchanged from the HPS
steady jet. For $x <Vt, \, t >0$, the jet will be the weak new jet
that was emitted from the central engine. Of course, an abrupt step
function change is a mathematical simplification, but the transition
region is much thinner than $L$. The change in emissivity is so
drastic that for $t< L/(2V)$, the older plasma at $x >Vt, \, t
>0$ is much more luminous than the newer plasma at $x <Vt, \, t >0$
and this dominates the integrated jet luminosity. Based on Figure 2
(and as verified in the analysis to follow) this describes the jet
evolution and emissivity for many hours. At late times, a new weak
jet may form and this is likely to have a more compact $\tau \sim 1$
surface since the number density and magnetic field are likely less.
However, this is after the decay and is not germane to the
discussion of the prompt decay of the 15 GHz flux density.

\begin{figure}
\begin{center}
\includegraphics[width=80 mm, angle= 0]{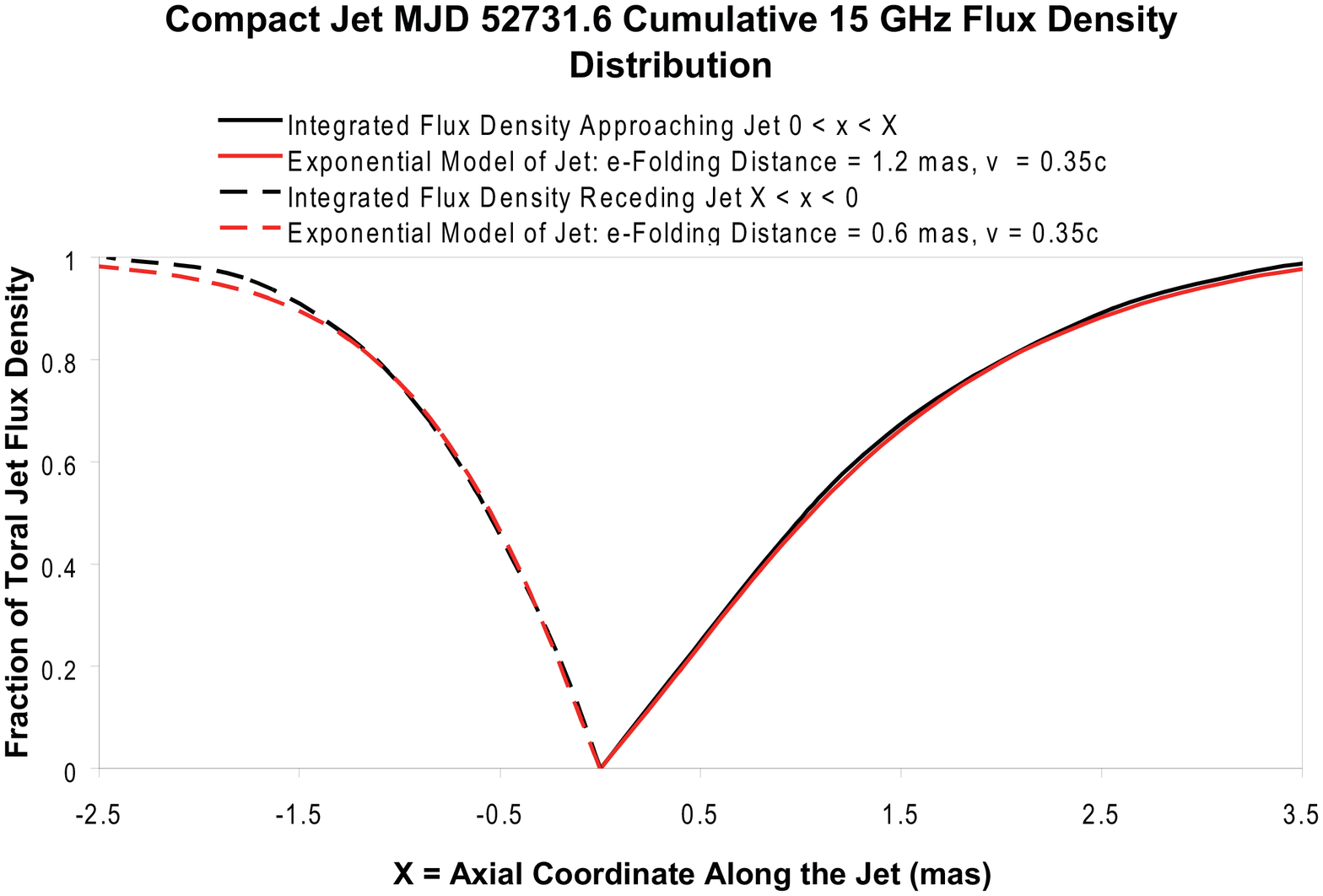}
\includegraphics[width=80 mm, angle= 0]{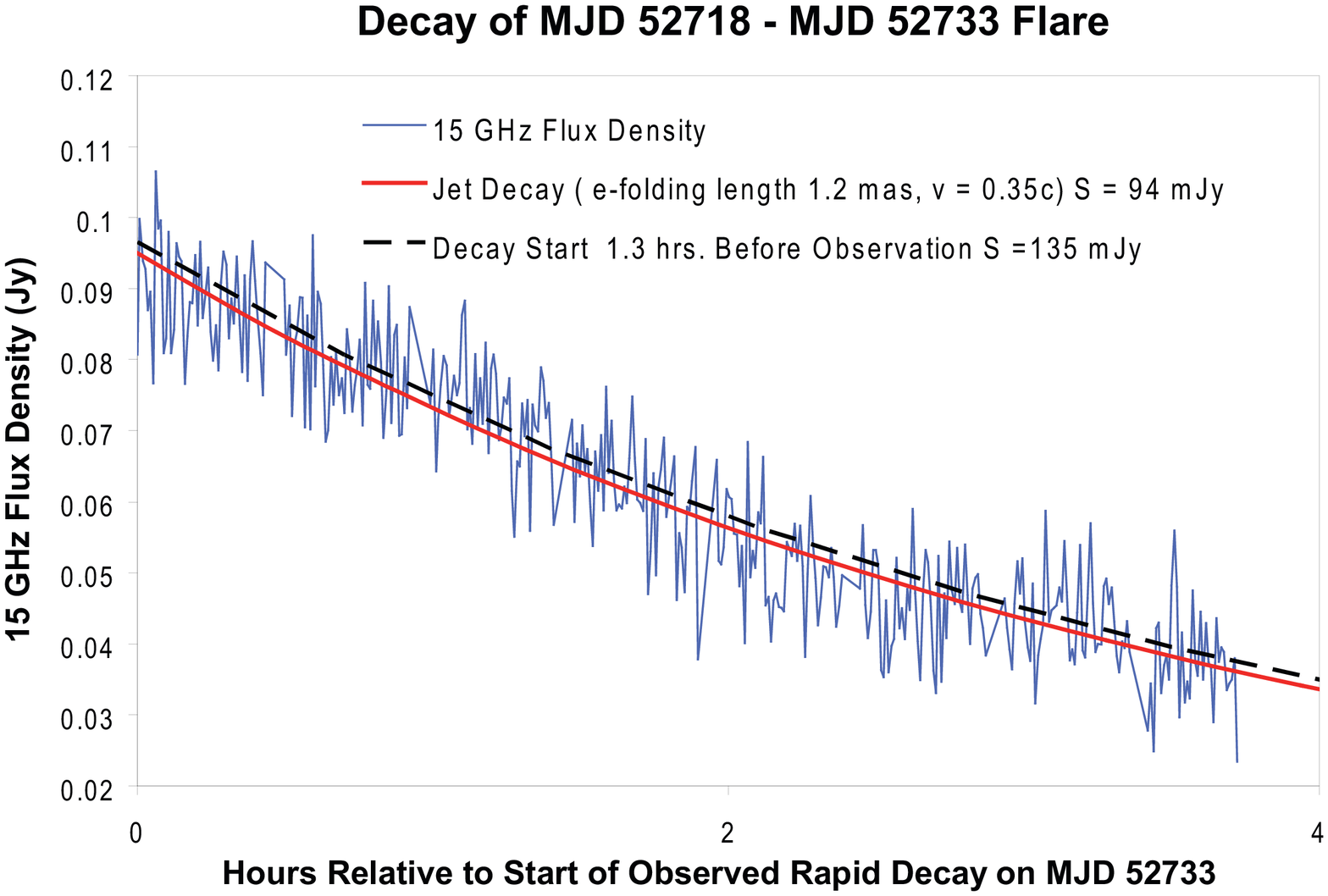} \caption{The solid
(dashed) black curve in the left frame is the cumulative 15 GHz flux
density along the approaching (receding) jet on MJD 52731.6. The red
solid (dashed) curve in the left hand frame is the output of the
exponentially decaying linear source model (described in Section
2.3) of the effective emissivity (i.e., the resultant emissivity
arising from the intrinsic emissivity, synchrotron self absorption
and Doppler beaming) of the approaching (receding) jet after
blurring by the finite bean width of VLBA and interstellar
scintillation. The intrinsic velocity of both jets is 0.35c. The
apparent approaching (receding) velocity is 0.36c (-0.25c). The
right hand frame shows the 15 GHz flux density as a function of time
for the decay of the steady state HPS jet displayed on the left hand
side, that was imaged 1.6 days earlier. The model of the decay is a
plot of Equation (13) that is based on the scenario in Figure 3. The
modeled jet decay is overlayed on the observed HPS decay from the
top frame of Figure 2. Note that there are two fits to the data. One
assumes the decay start was synchronous with the observation start
time (the solid red curve). The other assumes the decay started 1.3
hours earlier (the black dashed curve).}
\end{center}
\end{figure}
\par In summary, we are looking for a model of jet fading that is a
consequence of a drastic reduction in jet power and in general
leaves a weak jet behind. This simple model is a propagating
discontinuity that separates the strong jet region from the weak jet
region. Figure 3 describes our model. We emphasize that the simple
dynamic depicted has never been directly observed. However, it is
consistent with what we know about the compact jet from VLBA imagery
and radio light curves. In Figure 3, $t=0$ on the left indicates a
steady HPS jet, the type of jet that we will describe later in
Figures 4 and 5 and has been observed with VLBA \citep{fuc04,rib04}.
The 4 steps in the grey scale shading crudely indicate the decay of
luminosity along the jet length. The darker the grey, the brighter
the jet. The power source for the jet is suddenly diminished in
strength at $t=0$, except for a weak (light grey) jet that persists
during the event. For the sake of argument, the weak background
residual jet has an integrated 15 GHz flux density $\sim$10 mJy
while the HPS jet was originally $\sim$100 mJy. The high emissivity
plasma continues to propagate down the conduit formed by the steady
state HPS jet after $t=0$. The transition from light grey below to
darker grey above locates the position of the discontinuity in the
snapshots $\geq2$ hrs. As it propagates, the plasma upstream obeys
the same effective emissivity (the resultant emissivity that
includes the effects of Doppler beaming and SSA applied to the
intrinsic emissivity) profile as the HPS steady state jet in this
region as determined empirically from the VLBA radio images. The
original dark grey rectangle has moved downstream towards the top of
the figure and occupies the position of the fainter second rectangle
at $t = 2$ hrs. The next time step proceeds similarly. No new flow
of high emissivity jet material leaves the central engine, the
central engine for the jet is still nearly shutoff. The high
emissivity plasma that once occupied the lowest rectangle at $t=0$
continues advancing towards the top of the figure and occupies the
third position at $t = 4$ hrs. The jet luminosity decay continues
and this is represented by the lighter shade of grey.
\subsection{Previous Models of Compact Jets}
One approach to describing the jet spatial variation is based on
simple theoretical jet models. These were introduced into the
literature in \citet{bla79} which give some general features of the
effects of synchrotron self-absorption (SSA) and beaming in jets.
However, these power law models of the parameters were found to be
too simplistic to describe astrophysical jets and more elaborate
power law depictions of jets were created to deal with the
constraints imposed by observation \citep{ghi85,ghi89,ghi96}. These
models were developed for extragalactic radio sources. Similar types
of power law models for Galactic sources were applied to the jets in
Cygnus X-1 and MAXI J1659-152 \citep{kai06,par13}.
\par In the context of GRS~1915+105, models based on
\citet{ghi85,ghi89,ghi96} were applied to the compact jet
\citep{pun10}. The models described the surface brightness profile
of the compact jet considered here (in the next subsection and
Figure 4) as well as the broadband spectral energy distribution. In
order to satisfy all of the constraints of the observations, the jet
models were necessarily highly stratified in the transverse
direction. The inner most jet was the most collimated and the
magnetic field was turbulent in equipartition with the plasma
energy. The magnetic field became ordered outside of this core and
the jet expanded more rapidly with increasing transverse distance
from the core. The jet flared non-uniformly in the axial direction
as well. In the end, the parametric jet had 55 input parameters
describing the various power laws.
\par Power law models incorporate the intrinsic luminosity, SSA, Doppler beaming and
the basic dimensions to create an effective emissivity. The
complexity of the parametric model is not essential to understand in
the present context. In order to implement the model of the compact
jet fade in the last section, we are striving to understand the
spatial variation of the effective emissivity of the steady state
HPS jet. In this regard, it is not advantageous to assume that
actual jets are well described by numerous parameters that obey
power laws. This has no benefit in the present context and we strive
to find something directly related to the observations. Instead, we
choose to empirically determine the spatial variation of the source
of effective emissivity responsible for the images of the compact
jet in \citet{fuc04,rib04} during HPSs. Namely, we want to determine
the source function for the effective emissivity directly from the
radio images, independent of the theoretical model of the jet.

\subsection{An Empirical Source Function for the Compact Jet}

\par There have been two deep 15 GHz VLBA (milliarcsecond - resolution) observations of a HPS compact jet that have been published. The
first, was reported in \citet{fuc03,fuc04} and occurred on MJD
52731.6, 1.6 days before the HPS decay chronicled in the top frame
of Figure 2. The second occurred on MJD 52748.4 \citep{rib04}. We
note that compact jets were observed in \citet{dha00}, but these
were not HPS jets. We comment on these observations at the end of
this section. In order to estimate $D$, the total flux density as a
function of displacement along the approaching jet is calculated
from the radio image. The result of this integration for MJD 52731.6
is the black curve plotted in the left hand frame of Figure 4. This
is the ``steady" state before the decay. The curve is the cumulative
distribution of the detected 15 GHz flux density along the
approaching jet. It begins as 0\% at the radio core and equals 100\%
at the last contour in the radio image. The fact that this is not
the actual end of the jet is not significant since the flux density
is greatly reduced at these distances. The next step is to make a
comparison with a radio luminosity source function. In order to do
this we note three major facts
\begin{enumerate}
\item The jet is unresolved along its width so the source of
effective emissivity for the VLBA image can be represented by a line
source
\item It should be noted that this image is blurred (made longer) by the
finite beam-width of the interferometer and interstellar
scintillation. Based on the results of \citet{dha00}, it is
estimated that effective ``blurred beam-width" (that includes the
effects of interstellar scintillation) is 1.54 mas. The actual jet
is effectively viewed through this ``blurring aperture."
\item One other caveat is that the brightness of the jet within the vicinity of the
radio core depends on the jet velocity. The finite resolution of the
interferometer will detect emission from both the counter-jet and
the approaching jet in this region. The observed flux density,
$F_{\nu}\sim \nu^{-\alpha}$, in any beam is a Doppler enhancement of
the intrinsic flux, $F_{\mathrm{intrinsic}}$ \citep{lin85}:
\begin{equation}
F_{\nu} =\delta^{2+\alpha}F_{\mathrm{intrinsic}}\;,
\end{equation}
where
\begin{equation}
\delta= \frac{\sqrt{1-(v/c)^2}}{1-(v/c)\cos{\theta}}\;.
\end{equation}
Equation (2) provides an enhancement of the approaching jet flux and
a reduction of the receding jet flux. The net effect of Equation (2)
and the finite resolution of the restoring beam is to produce a
velocity dependent offset of the peak of the flux density of the
radio image from the true core position. The 1.54 mas effective beam
width centered on the peak flux will detect major fractions of both
the approaching and receding jets. The core shift is needed to
allocate proper amounts of the peak flux density to each side of the
jet. The fraction of core flux assigned to the approaching jet by
the implementing a core shift from the flux peak will be equal to
the fraction of the total jet flux in the approaching side - if
bilateral symmetry holds and the jet has a constant velocity. This
offset is required to identify a starting point for the cumulative
flux distribution of the approaching jet as observed with VLBA (the
black curve in the left frame of Figure 4). Yet, the resultant decay
length $D$ that arises at the end of the analysis (combined with the
decay time, $T$) must also produce the same velocity used in the
derivation of the offset. So the problem is nonlinear. For example,
in the analysis of this paper, the peak flux per beam is offset
downstream of the true core position by $\approx$ 0.3 mas. In order
to understand this core shift concept, and to appreciate the
nonlinearity, we verify that it is of the correct magnitude
explicitly after Equation (6).
\end{enumerate}

\par We are actually interested in the empirical source function for
the emissivity that was emitted by the source (not the image blurred
by scatter-broadening and the finite resolution of the VLBA) in
order to understand the length scale over which physical changes
take place. We have created such an empirical source function and
then mathematically blurred the image in order to replicate the
image made by VLBA. The resultant cumulative distribution of flux
density of the approaching jet is plotted as the red curve in the
left hand side of Figure 4. We can write the empirical source
function, $S(x)$, for the effective emissivity as
\begin{equation}
S(x)= (F_{0}/D)\rm{e}^{-x/D}\delta(v)^{2+\alpha}\Theta(x) +
(F_{0}/D_{rec})\rm{e}^{x/D_{rec}}\delta(-v)^{2+\alpha}\Theta(-x)\;,
\end{equation}
where $x$ is the actual axial displacement of the jet plasma from
the radio core and $\Theta(x)$ is the Heaviside step function. The
first term represents the approaching jet, $D$ is the e-folding
distance and $F_{0}$ is the normalization. The second term is the
source function of the receding jet. This is required because the
wide effective beam width of the observation samples the receding
jet as well near the origin of the approaching jet. We have
introduced $D_{rec}$, the e-folding distance of the linear source
function for the effective emissivity of the receding jet. In order
to transform this quantity to the observed emissivity,
$\mathcal{E}(y)$, one must convolve with a Gaussian beam. If y is
the observed displacement from the radio core,

\begin{equation}
\mathcal{E}(y)= \int_{-\infty}^{\infty}
{S(x)\frac{1}{\sqrt{2\pi}\sigma}\rm{e}^{\frac{-(x-y)^2}{2
\sigma^{2}}}\,dx} \;,
\end{equation}
where the standard deviation, $\sigma$, is the effective beamwidth
(1.54 mas) divided by 2.35 (the factor required to convert full
width at half maximum to standard deviation). The radio images were
made from a circular beam. The cumulative distribution,
$\mathcal{F}(y)$, that is plotted in red in Figure 4 is

\begin{equation}
\mathcal{F}(y)= \frac{\int_{0}^{y} {\mathcal{E}(z)\,dz}}
{\int_{0}^{\infty} {\mathcal{E}(z)\,dz}}\;.
\end{equation}

\begin{figure}
\begin{center}
\includegraphics[width=125 mm, angle= 0]{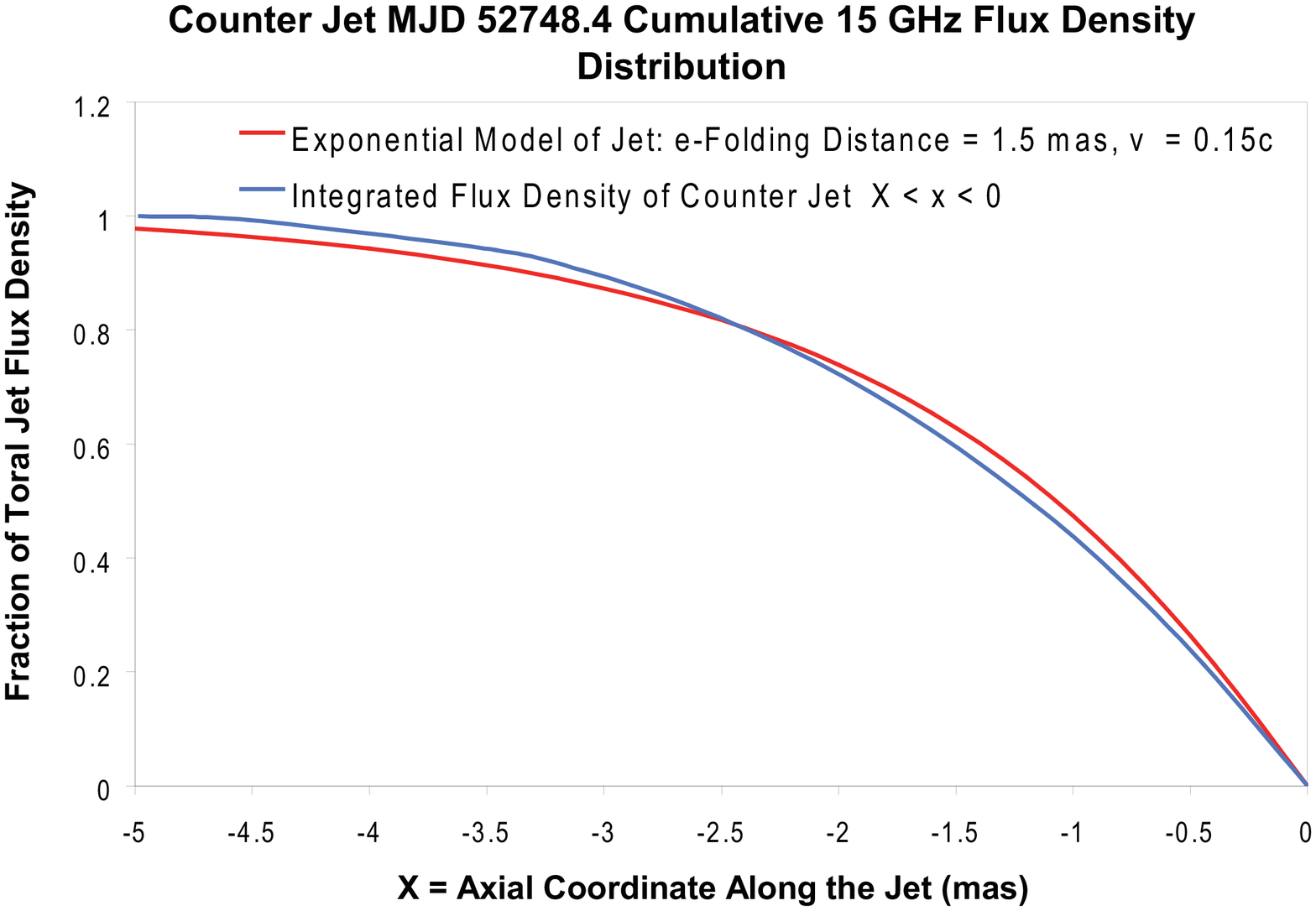}
\includegraphics[width=125 mm, angle= 0]{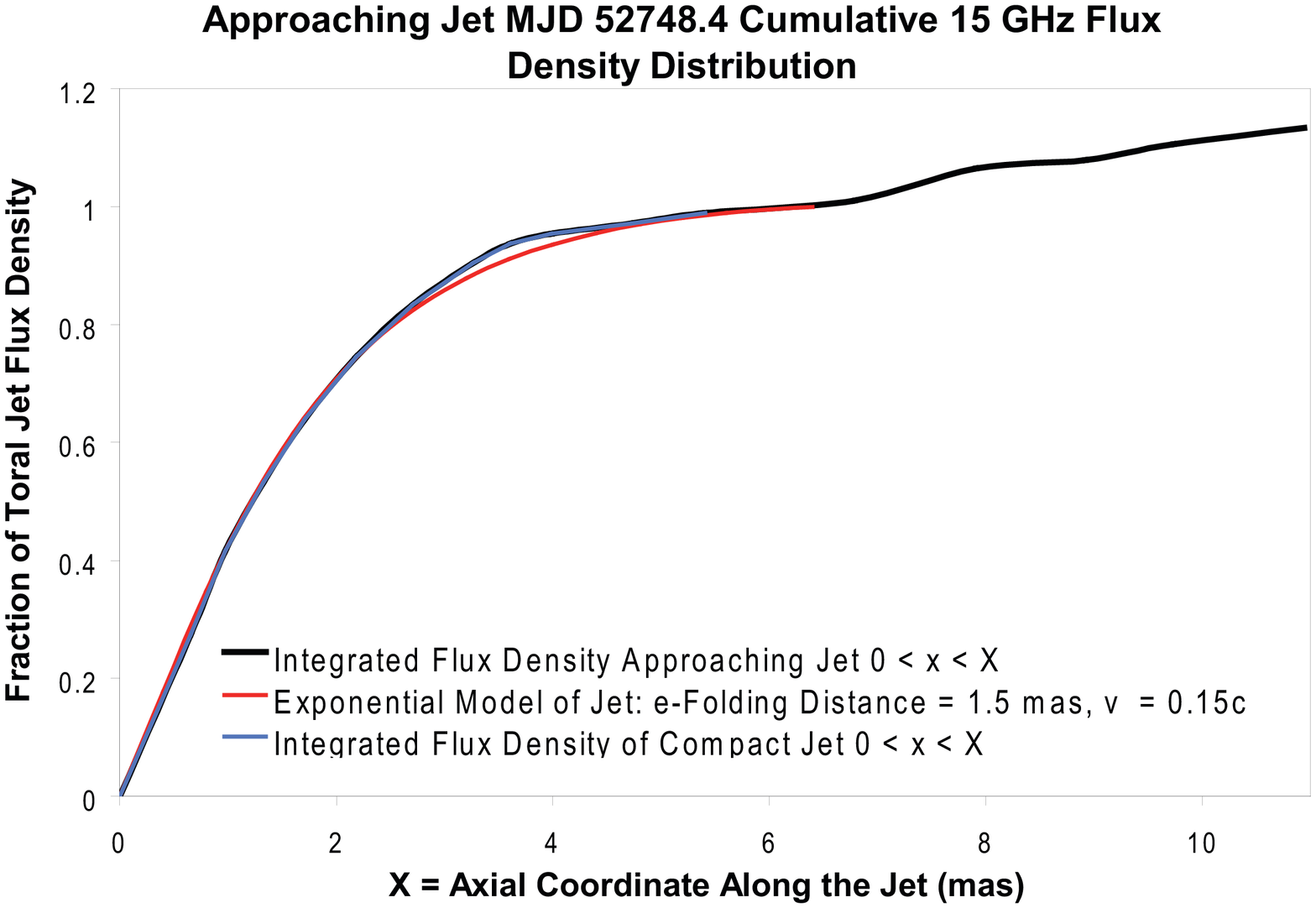}
\caption{The top frame is the fit of the empirical source of
effective emissivity of the counter jet on MJD 52748.4. The fit is
obtained simultaneously with the fit to the approaching jet in the
bottom frame. The comparison of the source to the VLBA image is
achieved through the application of Equations (4) - (10). The blue
curve represents the cumulative flux density of the compact jet in
the radio images. The red curve is the output of the exponentially
decaying linear source model of the jet effective emissivity after
blurring by the finite beam width of VLBA and interstellar
scintillation. The solid black line is the cumulative 15 GHz flux
density along the approaching jet on MJD 52748.4. Note the faint
discrete ejection 2-4 mas beyond the end of the compact jet. This
jet is significantly more symmetric than the jet in Figure 4
(especially if one segregates the discrete emission on the
approaching jet side). Consequently, the jet is consistent with a
slow intrinsic velocity of 0.15c.}
\end{center}
\end{figure}
\par At this point it is easier for the reader to appreciate the
core shift discussed in item (3) above. The empirical jet starting
point is ambiguous since this inherently asymmetric source is
sampled by a very wide effective beam (compared to the proper size
of the core). The effective beam (including interstellar
scintillation) is 1.54 mas (compare this to the e-folding distance
derived by this analysis of 1.2 mas in Figure 4). Thus, the beam is
large compared to the characteristic dimension of the jet. When the
beam is centered on the peak flux density ($F_{\nu}(\rm{peak})$)
location, it is capturing flux density from both the approaching
($F_{\nu}(\rm{core\;app})$) jet and the receding
($F_{\nu}(\rm{core\;rec})$) jet. We can write this as

\begin{equation}
F_{\nu}(\rm{peak}) = F_{\nu}(\rm{core\;app}) +
F_{\nu}(\rm{core\;rec})\;.
\end{equation}
We have two constraints. From Equation (2) and the assumption of
bilateral symmetry
\begin{equation}
\frac{F_{\nu}(\rm{core\;app})}{ F_{\nu}(\rm{core\;rec})} =
\frac{\delta(v)^{2+\alpha}} {\delta(-v)^{2+\alpha}} \;.
\end{equation}
We use the value of $\alpha =0.1 \pm 0.2$ for the jet on MJD 52731.6
that was previously determined from VLBA images \citep{rib04}. The
other constraint derives from the constant velocity assumed for the
jet which leads to the approximately constant Doppler enhancement
along the jet from Equations (2) and (3) if $\alpha \approx 0$, as
observed. Thus, the total jet integrated flux density fractions in
the approaching and receding jets should be approximately the same
as the ratio of flux density from the core assigned to the
approaching and receding jets

\begin{equation}
\frac{f}{1-f}\equiv \frac{F_{\nu}(\rm{total\; jet\;app})}{
F_{\nu}(\rm{total\;jet\;rec})} \approx
\frac{F_{\nu}(\rm{core\;app})}{ F_{\nu}(\rm{core\;rec})} =
\frac{\delta(v)^{2+\alpha}} {\delta(-v)^{2+\alpha}} \;.
\end{equation}
The next thing that needs to be computed is the effective core
offset in the radio image that assigns  a fraction, $f$, of the peak
flux density to the approaching jet and a fraction, $1-f$, of the
peak flux density to the receding jet. We need this offset in order
to know exactly where to start the integration of the cumulative
approaching jet flux density that is computed in the left hand side
of Figure 4. From item (1), a line source for the effective
emissivity is consistent with the data. Therefore, we can
approximate the offset, $\Delta$, of the start of the approaching
jet from the center of the beam located at the point of peak flux
density as
\begin{equation}
\Delta \approx \frac{F_{\nu}(\rm{core\;app})}{ F_{\nu}(\rm{peak})}
(1.54\,\rm{mas}) -1.54\,\rm{mas}/2 \approx (f-0.5)1.54\,\rm{mas} \;.
\end{equation}
As an example, if one were to split the core evenly with both the
receding jet and approaching jet then $f\approx 0.5$. The jet would
begin with zero offset, i.e. both the jet and counter jet start at
the point of peak flux density. By Equation (9), this corresponds to
$v\approx 0$. This linear approximation for the offset is only
accurate if the beam is much wider than the size of the magnitude of
the offset.

\par One can now appreciate how
nonlinear an iterative solution is in general. One has five
constraints, Equation (6) for both the jet and counter jet, the left
hand and right hand equalities for the flux ratio in Equation (9)
and Equation (10), and five unknowns $v$, $D$, $D_{rec}$, $f$ and
$\Delta$. Thus, a solution should exist if the model is a reasonable
representation of the physical situation. Trial values of $v$, $D$,
$D_{rec}$ and $\Delta$ must be chosen. The value of $\Delta$
determines the start point of the integration of the jet and counter
jet flux densities over their respective lengths. Using this
starting point, one can compute the total flux density in both the
approaching and receding jets, $F_{\nu}(\rm{core\;app})$ and
$F_{\nu}(\rm{total\;jet\;rec})$, respectively. From Equation (9)
this determines $f$ which in turn determines $\Delta$, by Equation
(10), hence the first aspect of the nonlinearity inherent to this
problem. The other aspect of nonlinearity of this problem is that
the chosen values of $v$ (the Doppler boosting of both the jet and
counter jet), $D$, $D_{rec}$ in conjunction with Equation (6) needs
to produce a cumulative flux density distribution that fits the
observed cumulative flux density distribution determined from the
VLBA observations. The observed cumulative flux density distribution
determined from the VLBA observations is not independent of $\Delta$
(i.e., $\Delta$ changes the origin of integration). So, one is
looking not just for an arbitrary solution for $f$ and $\Delta$, but
one that can be fit with the simple exponential empirical model.
Arbitrary values of $\Delta$ tend to produce observed cumulative
flux density distributions determined from the VLBA observations
with non monotonic curvatures which are poorly fit by the
exponential model. A third nonlinear aspect occurs when any of the
parameters $v$, $D$ or $D_{rec}$ are varied in the process of
finding a solution, the other two parameters must be changed in
order to fit the data due to the large beam size which captures both
the approaching and receding jets near the core. Many iterations are
required before agreement of the constraints are reached.

\par Note that the method above is not guaranteed to produce a
good fit if the empirical model is inherently inaccurate. However,
we have found some rather tight fits to the data. Our best fit is
shown on the left hand side of Figure 4. The parameters are $D=1.2$
mas for the approaching jet. This model also required $v =0.35c$,
which corresponds to an apparent velocity of the approaching jet of
$V_{app}=0.36$c. A similar fit to the receding jet yielded $D_{rec}=
0.6$ mas, consistent with the foreshortening of the receding jet
that is readily visible in the radio image. The velocity $v =0.35c$
corresponds to an apparent velocity of the receding jet of
$V_{app}=-0.25$c. The fraction of the flux in the approaching jet
was found self-consistently to be $f \approx 69\%$. This agrees with
the results of \citet{rib04}, but agreement need not occur in
general (see Figure 5 and the related discussion). The offset of the
peak flux relative to the start of the jet is $\Delta \approx
0.3\,\rm{mas}$.

\par
As an aside, we note that Equation (10) is not the same calculation
as the core offset due to SSA \citep{bla79}. We are not comparing
core positions at different frequencies as in a discussion of SSA
opacity. We are estimating the fraction of the peak flux in the very
wide beam (compared to the size of the jet) that is attributable to
the approaching jet at one frequency, 15 GHz. Certainly, there is a
relationship between the two ideas, and the offset estimate in
Equations (7) - (10) can be considered a crude independent method of
estimating the magnitude of the shift in position of the origin of
the jet relative to the peak of the flux density as is the intent of
SSA core shift analysis. This method requires
\begin{itemize}
\item a knowledge of the angle of the jet to the line of sight,
\item the velocity of the jet and counter jet ,
\item a detected jet and counter jet,
\item and a constant spectral index along the jet.
\item the effective beam width is comparable to the
spatial scale of variation (eg., $D$)
\item the effective beam width is much larger than the computed
offset
\item The jet and counter jet are unresolved in the transverse
direction
\end{itemize}
The method assumes that
\begin{itemize}
\item the velocity is constant along the detected jet and counter jet,
\item intrinsic bilateral symmetry,
\end{itemize}
In general, one does not have sufficient astrometric accuracy and
numerous high frequencies of sampling in order to accurately find
the core position by extrapolating the core shift to infinite
frequencies (in the SSA based method). Equations (7) - (10) can
provide a crude estimate of the core shift if the seven requirements
listed above are achieved.
\par
The computation of $V_{app}$ is based on $D$, the e-folding distance
of the linear source of effective emissivity. We now invoke the
model of the flare decay posited in Section 2.1 and illustrated in
Figure 3. This model is equivalent to identifying the time dependent
line source for the effective emissivity of the approaching jet,
$S_{\rm{app}}(x,\, t)$, after the central engine power has been
drastically reduced as

\begin{equation}
S_{\rm{app}}(x,\, t)= S_{0}(0.7/D)\rm{e}^{-x/D} \Theta (x- V_{app}t)
+ \epsilon (x)\Theta (x) \;,
\end{equation}
where $\Theta (x-V_{app}t) $ is the Heaviside step function and
$\epsilon (x)$ is the low level background residual jet (i.e., the
jet does not turn all the way off in general). $S_{0}$ is the
overall normalization. The coefficient of 0.7 results from the fact
that 70\% of the 15 GHz flux in the jet 1.6 days earlier (on MJD
52731.6, the left hand frame of Figure 4) is in the approaching jet
and 30\% in the receding jet \citep{rib04}. We assume this same
fraction. The light curves in Figure 2 represents the total
integrated luminosity. Therefore, we need the contribution from the
receding jet as well. For the receding jet, one has
\begin{equation}
S_{\rm{rec}}(x,\, t)= S_{0}(0.3/D_{rec})\rm{e}^{x/D_{rec}} \Theta
(-x + V_{rec}t) + \epsilon_{\rm{rec}}(x)\Theta (-x) \;.
\end{equation}
Assuming bilateral symmetry in Equation (1) and a line of sight of
$60^{\circ}$ (i.e., $-120^{\circ}$ for the receding jet) determined
in \citet{rei14}, the receding apparent velocity is $V_{rec}$
=-0.25c. At any time $t$, after the jet central engine was reduced
in power, the total flux density, $F_{\nu}$, emitted from the
compact jets is
\begin{equation}
F_{\nu}(t) = \int_{-\infty}^{\infty}{S_{\rm{app}}(x,\, t) +
S_{\rm{rec}}(x,\, t)}\,dx \approx 0.7S_{0}\rm{e}^{-V_{app}t/D} +
0.3S_{0}\rm{e}^{V_{rec}t/D_{rec}} \;.
\end{equation}
The exponential line source from Equation (4) for the effective
emissivity with an e-folding distance for the approaching (receding)
jet of $D =1.2$ mas ($D_{rec}= 0.6$ mas) and a bulk velocity
$V_{app}$ = 0.36c applied to Equations (11) - (13) can be used to
quantify the simple fading jet model of Section 2.1. Equation (13)
yields the fit to the decay data presented in Figure 2 in the right
hand side of Figure 4.  We plot the curve two ways in order to
investigate the affect of shifting the origin of time. The solid red
curve is a plot of the decay if it started when the observation
started. The overall normalization is $S_{0} = 94$ mJy. The heavy
dashed black curve has a starting flux density of 135 mJy when the
decay started, similar to the level the day before (see Figure 2).
This requires that the observation started 1.2 - 1.4 hrs before the
observation in order to achieve a good fit. The decay is not that
sensitive to the start time for an exponential decay as can be seen
from Equation (13). The larger term from the approaching jet is also
from the longer jet with the larger decay time, so this larger
exponential function dominates more as time elapses and the decay
gets closer to a pure exponential. There is a slight excess at late
times to the fit in the right hand side of Figure 4, this might
represent emission from the weak new jet. In summary, this empirical
source function for the effective emissivity reproduces the compact
jet in the HPS on MJD 52731.6 and (in the context of the simple jet
fading model) the subsequent decay on MJD 52733 just 1.6 days after
the jet was imaged.
\par Even though a major advantage of studying HPS is to provide a
stable state to analyze, it should be noted that in general 1.6 days
is a long time interval for dynamic change in GRS~1915+105. Major
changes can occur in the X-ray and radio on time scales of seconds
or minutes \citep{kle02}. Although, we have no evidence that this
occurs in HPS, it might not be valid to assume that the $D$ value
determined in Figure 4 typified the jet just prior to the decay.
Without simultaneous data, this can only be estimated by studying
multiple VLBA images of HPS. This is addressed below.

\par The most sensitive radio imaging of an HPS jet was on MJD
52748.4 \citep{rib04}. Figure 5 shows the fit of a linear decaying
source of the effective emissivity to the cumulative flux
distribution along both the approaching and receding jets. The fit
is defined by $D=D_{rec}$ = 1.5 mas and $v$ =0.15c. This jet is much
more symmetric than the jet on MJD 52731.6 and therefore requires a
much smaller bulk velocity. The decay length is slightly larger than
that on MJD 52731.6 in Figure 4. There is one interesting feature
that is unique to this radio image. A weak discrete ejection appears
from 2 to 4 mas beyond the end of the compact jet. Comparing with
the 8.3 GHz radio image in \citet{rib04} indicates that it is
optically thick (flat spectrum). The feature is 10 times the rms
noise in the image and seems to be real, but without a deeper image
showing a similar feature we cannot rule out this as an artifact of
inadequate. u-v coverage. Thus, we must consider this a tentative
detection of a discrete component. A third unpublished HPS compact
jet observation with comparable u-v coverage and sensitivity to the
two images analyzed in this paper was also
performed\footnote{\citet{rib04} unpublished result presented at the
EVN Symposium 2004 The 7th European VLBI Network Symposium on New
Developments in VLBI Science and Technology Toledo, Spain, 2004
October 12-15)} and occurred on MJD 52722 and is also consistent
with $D \approx$ 1.2 - 1.5 mas.

\par The range of 1.2 mas $<D<$ 1.5 mas is consistent with the two
other 15 GHz VLBA published observations of compact jets with
comparable u-v coverage, flux density and sensitivity on MJD 50914
and MJD 50935 \citep{dha00}. However, these jets were not produced
during a HPS. It is worth noting that the much weaker (45 mJy)
compact jet observed on MJD 50744 reported in \citet{dha00} was fit
with an e-folding distance of $<1$ mas. This small value might be an
artifact of the insufficient sensitivity to detect the faint jet
extremities.
\par The 1.6 days between the VLBA observation on MJD 52731.6 and
the decay on MJD 52733, is quite long for GRS~1915+105, so one
cannot assume that this was the jet length during the decay as noted
above. From Figure 2, the e-folding time is bound by the existing
observations by 3.8 hrs $<T<$ 8.6 hrs. From Figures 4 and 5 and the
discussion that followed 1.2 mas $<D<$ 1.5 mas. Then from Equation
(1), this creates a range of apparent jet velocities, $0.16c <
V_{app}< 0.46c$ and a range of intrinsic jet velocities of $0.17c <
v< 0.43c$. The high end of the range is corroborated by other data.
It is driven by the rapid decay on MJD 52733 of only 3.6 hrs. It is
very likely that this rapid decay is the result of this being a high
velocity jet state. It was noted in \citet{rib04} that this was the
most asymmetric compact jet ever detected. Thus, their analysis of
jet speed from Doppler asymmetry produced the largest $v$ of any
compact jet by a considerable amount. Correcting their result for
the new line of sight (see the Discussion) from the parallax
measurements of \citet{rei14}, yields $v =0.37 \pm 0.04$c, in
agreement with the maximum estimated $v$ attained in this study.
Another argument for this being the high end of the velocity range
is that Table 1 indicates that this HPS was much more luminous than
any other observed HPS.
\subsection{Comparison to Alternative Models of Jet Fading}
A key assumption of this study is the scenario for jet fading
described in Section 2.1 and illustrated schematically in Figure 3.
In this scenario, the jet fading is a transient decay of a strong
jet that was abruptly suppressed at its source. There is no direct
observational evidence to support this idea and the scenario is not
unique. Alternatively, one might think of the jet fading as a slow
turnoff of the jet source in which one of the steady state power law
solutions of \citet{pun10} is slowly varied and no strong transient
or contact discontinuity is created. The lone restriction here is
that the plasma from the full power jet episode must propagate
sufficiently far down the jet so that its contribution to the total
flux density is negligible after an e-folding time $T$, i.e $V_{app}
\gg D/T \approx 0.36c$. If this were not true there would have to be
a conspiratory connection between turning down the power source at
the base of the jet and the rate that previously ejected high
emissivity plasma is losing its surface brightness in order to
produce an exponential decay in time - not only once but on both MJD
52733 and on MJD 53122.
\par In order to assess the plausibility of these ideas, we first
note the nontrivial successes of the jet fading scenario presented
here:

\begin{enumerate}
\item{The model provides a physical connection between the
spatial exponential decay profile of the source of effective
emissivity of the steady state HPS jet and the temporal exponential
decay of the fading of the jet emission 1.6 days later}
\item{The constraint imposed by the above physical connection
implies a jet velocity that agrees very closely with the jet
velocity found in \citet{rib04} by considering jet bilateral
asymmetry due to Doppler abberation of an intrinsically bilaterally
symmetric system}
\end{enumerate}

\par By contrast, consider these issues in the context of
a slowly varying jet scenario.

\begin{enumerate}
\item{The spatial exponential decay profile of the source of effective
emissivity of the steady state HPS jet and the temporal exponential
decay of the fading of the jet emission 1.6 days later are
coincidental. The temporal decay profile is a manifestation of how
the local physics at the base of the jet alters the parameters
injected into the jet model. It is coincidental that this follows an
exponential profile on both MJD 52733 and on MJD 53122. }
\item{The constraint, $V_{app} \gg D/T \approx 0.36c$ noted above (that is required to produce the temporal exponential
decay) disagrees with the jet velocity found in \citet{rib04} by
considering jet bilateral asymmetry due to Doppler abberation of an
intrinsically bilaterally symmetric system}
\end{enumerate}
One other curious feature of the slowly varying jet scenario is
evident if the details of the power law models of \citet{pun10} are
considered. The base of the jet that produces the majority of the 15
GHz flux density is located $<3 \times 10^{7}$ cm from the central
black hole (see Table 1 and Figure 2 of \cite{pun10}). Considering
the time scales for decay of 3.8 hrs and 8.6 hrs from Figure 2,
turning down the power of the jet by moving plasma in the disk at
the base of the jet corresponds to a rather suspect slow radial
velocity of $\sim 10^{3}$cm/s.
\par In summary, the transient model implies one value of $V_{app}$ that is consistent with the cumulative flux
distribution, the rate of compact jet decay and the bilateral
asymmetry. The slowly varying jet model is not consistent with
bilateral asymmetry and the variations in time and space are
unrelated observable parameters. We choose to consider the
interpretation that has descriptive power and is more consistent
with other observations, although we can not rule out the slowly
varying jet model.
\section{X-ray Properties of the Decay of High Plateau States}
The decay of the HPS on MJD 53122 in Figure 2 was accompanied by
Rossi X-ray Timing Explorer (RXTE) observations. Both PCA
(Proportional Counter Array) and HEXTE (High Energy X-ray Timing
Experiment) observations were performed on MJD 53122.107 and again
on MJD 53122.179. For the sake of exploring the time evolution of
the disk - jet system, we reduced these data plus the data from the
height of the HPS 5 days earlier on MJD 53117.114.
\subsection{Observations and Data Reduction}
In order to characterize the source X-ray behavior during the HPS,
we looked at data that were simultaneous with the HPS that is best
followed at radio frequency. We reduced and analyzed Rossi X-ray
Timing Explorer (RXTE)/Proportional Counter Array (PCA) data taken
during observations that began on MJD 53100.230, 53117.114,
53122.107, 53122.179.
\par The data were reduced in a very standard manner (as in e.g.\citet{rod09}) with the
{\tt{HEASOFT}} V6.16 software suite. Energy spectra were obtained
from the top layer of the Proportional Counter Unit (PCU) \#2,
background spectra were produced from the ``bright source" model,
and responses produced with {\tt{pcarsp}} for the appropriate PCU,
layer and anode. We added 0.6\% systematic uncertainties to all
spectral channels, and we analyzed the resultant spectra within
{\tt{XSPEC}} v12.9.0.
\par
High resolution ($7.8125\times10^{-3}$~s) light curves covering the
full PCA spectral range were obtained from different data modes in
different observations. On MJD 53100.230, we used the {\tt{GOOD
XENON}} data, while for the other three observations we combined
{\tt{BINNED}} and {\tt{EVENT}} that respectively cover the low
(0--35 or $\sim 2-15$ keV) and high (36--255, $>15$ keV) spectral
channels.  PDS from each individual observation were produced with
{\tt{POWSPEC}} on intervals of 64~s, all intervals were further
averaged.

\begin{figure}
\begin{center}
\includegraphics[width=120 mm, angle= 0]{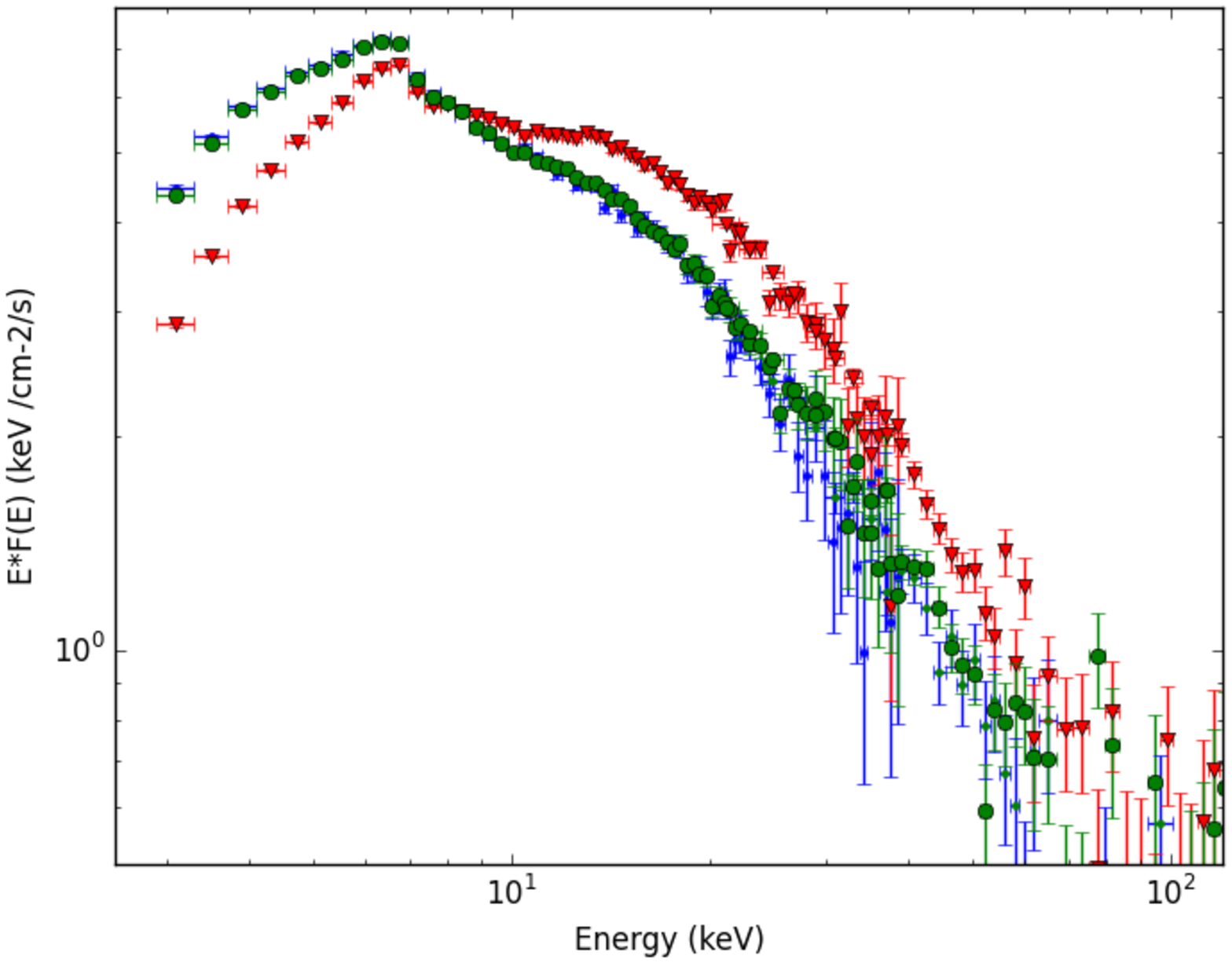}
\includegraphics[width=120 mm, angle= 0]{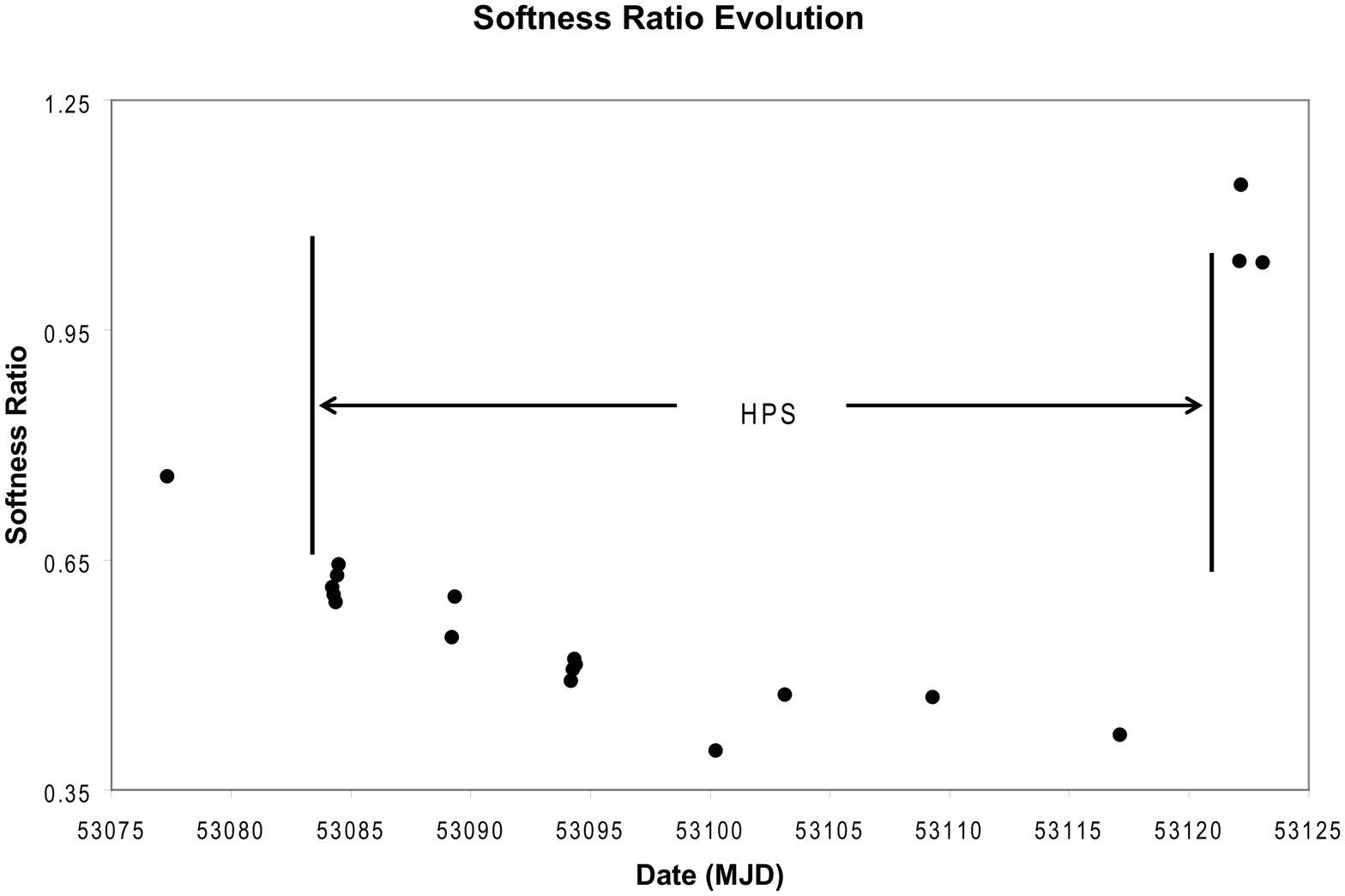}\caption{The top frame shows the X-ray SEDs
during an HPS and during the decay. The red plot is from MJD
53117.114, green from MJD 53122.107 and blue from MJD 53122.179. The
SED is much softer during the decay of the HPS state than during the
HPS. The bottom frame shows the time evolution of the softness
ratio, $R$, during the HPS and the decay of the HPS: $R=
F_{1.5-3}/F_{12-50}$, where $F_{1.5-3}$ is the intrinsic flux in the
interval 1.5 keV - 3 keV and $F_{12-50}$ is the flux in the interval
12 keV - 50 keV.}
\end{center}
\end{figure}
\begin{figure}
\begin{center}
\includegraphics[width=120 mm, angle= 0]{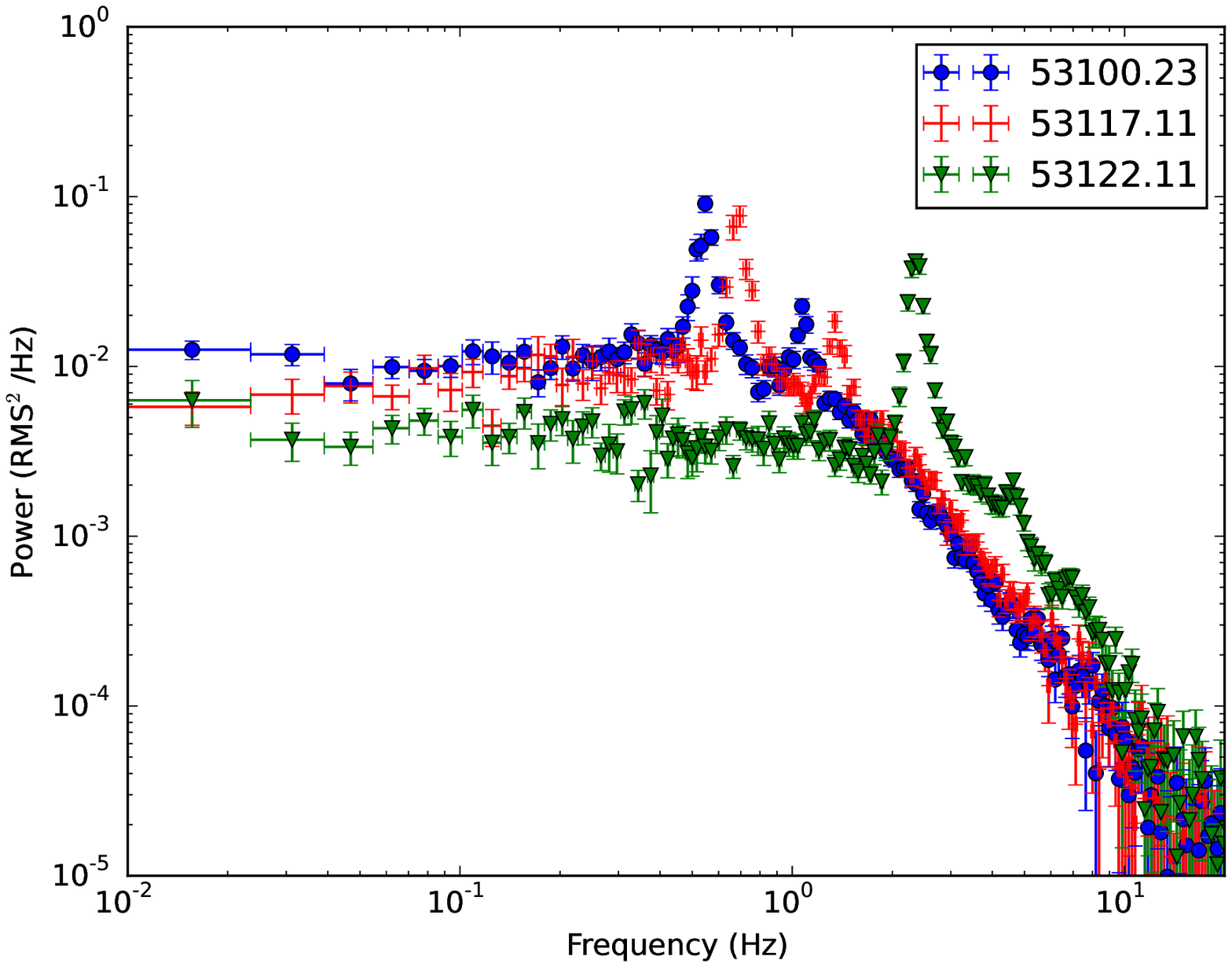}\caption{The Figure compares the PSD
during the HPS. Both the PSD on MJD 53100.230 in blue on MJD
53117.114 in red are from the HPS. The PSD of the combined data from
MJD 53122.107 and MJD 53122.179 is plotted in in black.}
\end{center}
\end{figure}

\subsection{Results} In order to see the spectral evolution, we show a plot of
the spectral energy distributions (SEDs) in the top frame of Figure
6. The first epoch is MJD 53117.114 in red, which occurs during a
HPS (see Figure 1). It is a much harder spectrum than the two
spectra obtained on MJD 53122.107 (green) and MJD 53122.179 (blue)
during the decay of the same HPS (see Figure 2). The plot shows that
the spectral softening appears to continue between MJD 53122.107 and
MJD 53122.179.

\begin{table*}
 \centering
\caption{X-ray Spectral Fits} {\tiny
\begin{tabular}{ccccccccccc}
 \hline
1                   & 2         & 3         & 4       & 5       & 6     & 7     & 8     & 9 & 10 & 11 \\
Epoch               & Line      & Line      & Line   & E        & E     & Power & Power & Flux & Flux & Reduced \\
Date                &  Center    &  Sigma   & Norm    & Cutoff   & Fold  &  Law & Law    & $10^{-8}$ & Intrinsic & $\chi^{2}$\\
(MJD)               &   keV      &  keV      &         &  keV     & keV   &    1  & 2     & $\mathrm{ergs/s/cm^{2}}$& $10^{-8}\mathrm{ergs/s/cm^{2}}$& (dof)\\

\hline
53117.114           &  $5.86\pm 0.10$    & $1.61\pm 0.08$ & $0.13\pm 0.01$  & $13.9\pm 0.2$    &  $25.4\pm 1.0$ & $2.11\pm 0.06$ &$0.54\pm 0.69$ & $2.12\pm 0.03$ & $3.15\pm 0.05$ & 1.46(100)\\

53122.107           &  $5.69\pm 0.27$    & $1.24\pm 0.10$ & $0.08\pm 0.01$  & $14.3\pm 0.4$    &  $34.9\pm 2.9$ & $2.51\pm 0.02$ & ...           & $2.14 \pm 0.03$ & $3.80\pm 0.06$ & 1.00(96)\\

53122.179           &  $6.29\pm 0.13$    & $1.10\pm 0.10$ & $0.08\pm 0.01$  & $14.7\pm 0.4$    &  $35.2\pm 1.8$ & $2.55\pm 0.02$ & ...           & $2.14 \pm 0.03$ & $3.80\pm 0.06$ & 1.30(87)\\

\hline
\end{tabular}}
\end{table*}

\par The spectral evolution can be formally characterized by
parametric spectral fits. We tried models that required an accretion
disk, to look for signs of spectral evolution, but there is no
evidence of a disk in any of the data. Our model had the following
parameters. The absorbing column of neutral hydrogen, $N_{H}$, was
allowed to vary for MJD 53122.107 and MJD 53122.179, in the models
described in Table 2, we found values of $5.69 \pm 0.27 \times
10^{22} \mathrm{cm}^{-2}$ and $5.88 \pm 0.26 \times 10^{22}
\mathrm{cm}^{-2}$, respectively. Since the values were similar, we
fixed $N_{H}= 5.7 \times 10^{22} \mathrm{cm}^{-2}$ for MJD 53117.114
which constrained the power law better in the more complicated fit
for this epoch in Table 2. We note that if $N_{H}$ was left free in
the fit on MJD 53117.114, it optimized at $N_{H}= 6.3 \times 10^{22}
\mathrm{cm}^{-2}$. Columns (2) - (4) of Table 2 describe the
properties of the Gaussian line used in the fit. Columns (5) and (6)
are the cutoff energy and e-folding energy required by the fit,
respectively. The particle number density power law spectral index,
$\Gamma$, is tabulated in Column (7). The epoch MJD 53117.114,
during the HPS, has a high energy excess and requires a second power
law that is poorly constrained by the data, as is indicated by the
large relative uncertainty in column (8). The hard second power law
comprises only 1.5\% of the flux from 1.5 - 150 keV (extrapolated
down to 1.5 keV). The total observed flux from 1.5 - 150 keV
(extrapolated down to 1.5 keV) is listed in column (9). The
intrinsic flux (corrected for $N_{H}$) from 1.5 - 150 keV is
tabulated in column (10). The reduced $\chi^{2}$ and degrees of
freedom of the model is in the last column. The most striking
evolution of the X-ray spectrum from the fully powered up jet state
to the jet decay is provided by the spectral index of the underlying
power law in column (7). The power law steepens from $\Gamma = 2.11$
to $\Gamma = 2.55$. Perhaps the best way to characterize the
steepening of the X-ray spectrum during the jet decay is by looking
at the intrinsic flux ratio, $R= F_{1.5-3}/F_{12-50}$, where
$F_{1.5-3}$ is the intrinsic flux in the interval 1.5 keV - 3 keV
and $F_{12-50}$ is the flux in the interval 12 keV - 50 keV. The
fluxes were computed from PCA observations with the assumption (in
contrast to Table 2) that $N_{H}= 5.7 \times 10^{22}
\mathrm{cm}^{-2}$ and models otherwise similar to those described in
Table 2. The fluxes below 2.5 keV were extrapolated from the model
and would not accurately depict any low energy contribution from a
thermal disk component. These data reductions and extrapolations
were originally performed in \citet{pun11} as part of a method that
calibrated counts from the RXTE All Sky Monitor with PCA determined
fluxes. On MJD 53117.114, $R=0.42$ compared to $R=1.14$ on MJD
53122.179. This softness ratio, $R$, is plotted during the course of
the HPS and during its decay in the bottom frame of Figure 6. There
is a dramatic change during the decay. Considering the top frame of
Figure 6, this affect is driven primarily by a change in the soft
X-ray flux. The physical interpretation would seem to be that there
are less low energy Comptonizing electrons in the jet state and more
low energy Comptoninzing electrons in the decay state.
\par Figure 7 contains plots of the power spectral density
(PSD) on MJD 53117.114 in red with the PSD of the combined data from
MJD 53122.107 and  MJD 53122.179 in black. Energy in the temporal
variations migrates from lower frequency during the HPS to
significantly higher frequency during the jet decay. As a reflection
of this, the fundamental quasi-periodic oscillation (QPO) frequency
changes by a factor of 3.5, from 0.69 Hz to 2.43 Hz. We also plot
another PSD during the HPS, on MJD 53100.230. During the entire HPS
the QPO fundamental frequency is less than 1 Hz and this frequency
dramatically changes during the decay as noted above.
\section{Short Time Scale Variability of Radio Emission} Previously, short time scale
variability of the 15 GHz flux density has been associated with
compact jets, but not during X-ray classes of type $\chi$,
equivalent to ``hard" X-ray spectra \citep{poo97,kle02,mir98}. The
associated time scales were from 10 - 45 minutes. It is formally
difficult to assess the spectral index of the radio emission,
because there is a lag between low frequency and high frequency
emission cycles. The spectra of strong quasi-periodic radio emission
are typically flat or inverted giving the appearance of synchrotron
self-absorbed spectra, but at certain times in the cycle the
spectrum can appear to be optically thin and steep
\citep{mir98,dha00,fen00}. Even the stronger periodic emission could
be a superposition of ``bubbles" at different stages of their
adiabatic decay. Each bubble begins flat spectrum and becomes steep
spectrum after adiabatic expansion \citep{van66}. The important
aspect to be noted here is that some of these states have been shown
to correspond to a compact jet with VLBA radio images that
morphologically appear similar to the jets in HPS \citep{dha00}.
Much weaker levels of radio emission have also been found to vary on
much shorter time scales \citep{pra10,rod09}. Even though no
quasi-periodic radio emission has been found in $\chi$ states, the
similarity in the radio images between jets in HPS and from
quasi-periodic events shown in \citet{dha00} suggests that there is
a connection between these phenomena. Likewise, we are interested in
the conjecture that the VLBA images of compact jets are deceiving
due to insufficient u-v coverage and sensitivity and there are
numerous ghost features moving quickly along the jet that are not
imaged \citep{dha00}. If so, there might evidence of these in the
time domain. For the quasi-periodic events that were imaged by VLBA
that looked similar to the compact jets there is clear evidence of
characteristic time scales in the time domain \citep{dha00}. Thusly
motivated, a search for rapid variations in HPS was initiated in
order to see these time scales are also generic time scale of
variation for compact jets.
\par First, the magnitude of the fluctuations are quantified. The
preliminary step was to compute a linear fit to each of the 95 Ryle
observations (those plotted in Figure 1) of the 8 HPS. A residual
from the linear fit was computed every instance of data capture,
typically every $\approx 32$ s. The absolute values of the residuals
to the linear fits were compiled as the distribution in the left
hand frame of Figure 8. The distribution of these fluctuations is
well fit by a Gaussian function of mean zero (plotted as a solid red
curve) with a standard deviation of 7.7 mJy. The quoted instrumental
uncertainty is $\approx 6$ mJy for the Ryle telescope with 32s
integration \citep{poo97}. Thus, there is possibly a slight excess
in these states over the systematic instrumental rms. These are not
large fluctuations. One can interpret the residuals from the linear
fits to the individual Ryle observations as being distributed as a
standard normal distribution. With this interpretation, a simple
convolution of standard normal distributions indicates that the data
is consistent with a standard normal distribution of intrinsic
fluctuations with a standard deviation $\lesssim 5$ mJy, i.e.,
$\sigma^{2} =\sigma_{\mathrm{systematic}}^{2} +
\sigma_{\mathrm{intrinsic}}^{2} =6^{2} + 4.8^{2} =7.7^{2}$.
Alternatively, since these are highly luminous states, the
systematic error might just be larger. Figure 8 does not show any
excess in the tail of the normal distribution. There is no evidence
for the large (20 mJy -50 mJy) oscillations that were discussed in
\citet{poo97,kle02,mir98} for quasi-periodic states.
\par Secondly, there are 15 observations of over 3 hrs of the HPSs. These are
combined to produce the power spectrum in the right hand frame of
Figure 8. The power spectrum was computed as the amplitude of the
complex Fast Fourier Transform of the auto-correlation function of
the ensemble. The normalization is chosen so that the integral of
the power over the sampled frequency range ($5.49 \times 10^{-5}$ Hz
to $ 1.40 \times 10^{-2}$ Hz) is equal to unity (i.e., the integral
of the PSD equals the total power in the fluctuations). In order to
interpret the spectrum, we compare it to simple models of noise. To
make a comparison to the Fast Fourier Transform result, we compute
the complex Fourier integral of the autocorrelation function over
positive frequencies then take the amplitude. The power spectrum is
well fit by the power spectrum of red noise with a break to white
noise at higher frequency \citep{kit61}. To facilitate the
comparison requires re-normalizing the red noise spectrum, with a
white noise break at high frequency, so that the integral of the
composite theoretical power spectrum from $5.49 \times 10^{-5}$ Hz
to $ 1.40 \times 10^{-2}$ Hz is unity, the same as the normalization
of the ensemble PSD. The fact that the curvature of the red noise
spectrum is not captured by the ensemble PSD, does not allow us to
create a strong constraint on the characteristic correlation time of
the red noise. The only constraint that we have is that the
correlation time is $>3000$ s. The fit at low frequency is improved
for longer correlation times. The PSD transitions from red noise to
white noise at frequencies above 0.0056 Hz. There is no excess power
associated with the time scales reported for the quasi-periodic
radio states in \citet{poo97,kle02,mir98}. Based on the variability
analysis, the compact jets in the HPSs are a distinct phenomenon
from quasi-periodic radio events. Whatever the cause of the
quasi-periodic variations, these variations are suppressed by the
strong compact jet associated with the HPS. There is also no
evidence of ghost features in the VLBA images that might be moving
with a much larger bulk velocity than was found here. The time
domain analysis illustrated in Figure 8 indicates a continuous jet
structure with minimal contribution from transient features.
\begin{figure}
\begin{center}
\includegraphics[width=80 mm, angle= 0]{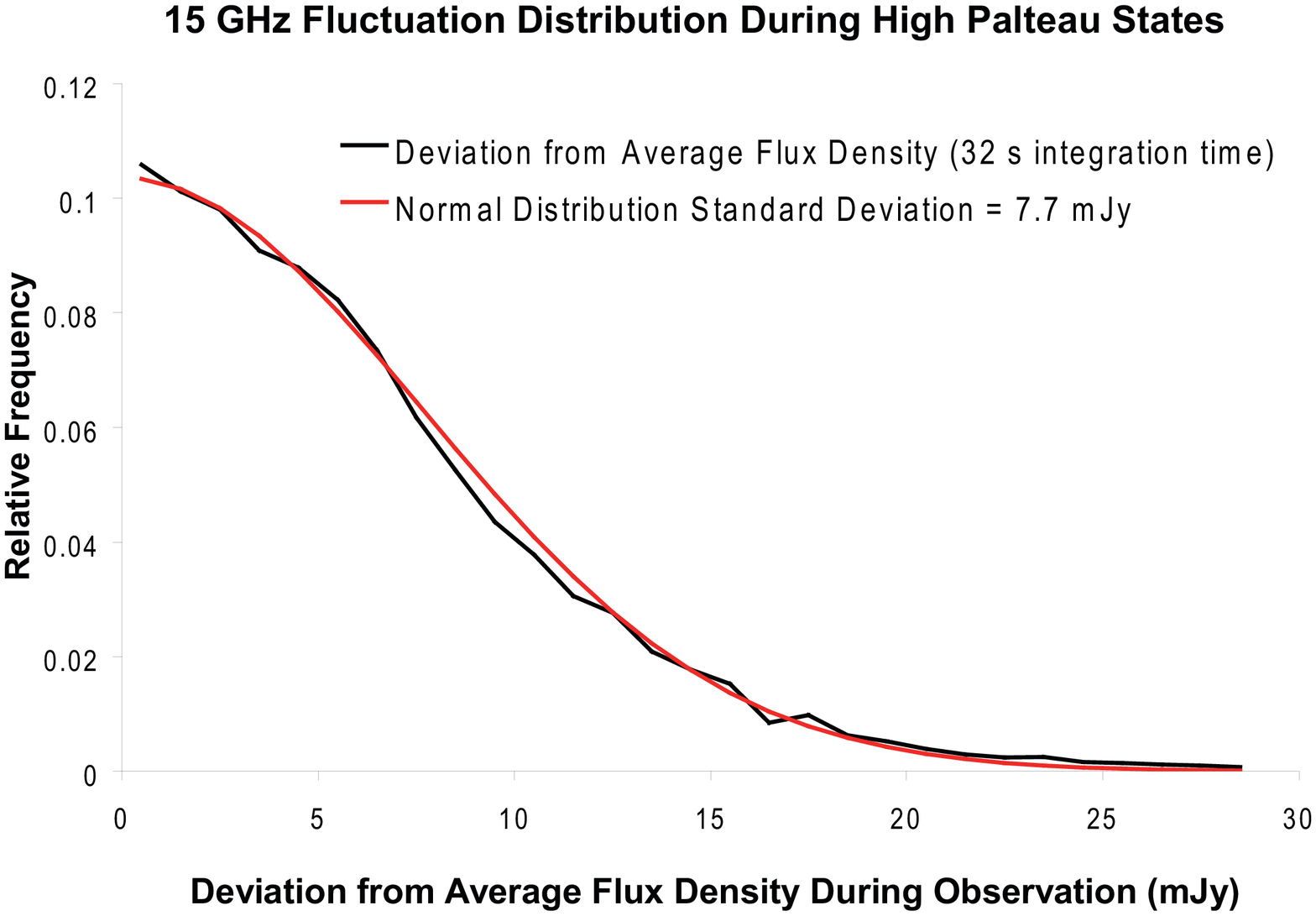}
\includegraphics[width=80 mm, angle= 0]{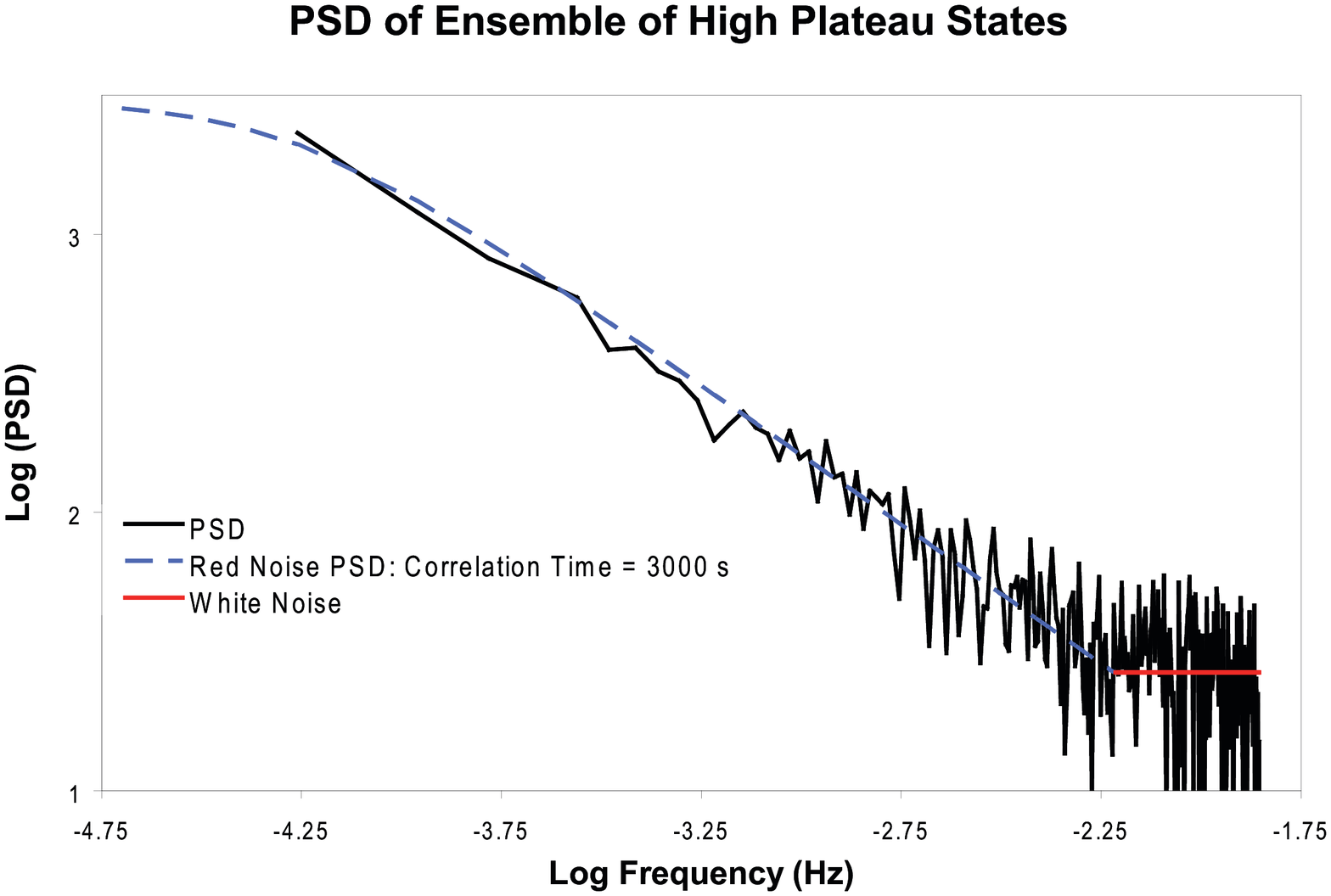} \caption{The solid
red line in the left hand frame is the Gaussian fit to the amplitude
spectrum of fluctuations from the ensemble of HPS plotted in black.
The right hand frame is the power spectrum of the ensemble of
fluctuations. A theoretical red noise power spectrum (with a break
to white noise at high frequency) that fits the data closely is
plotted as a blue dashed line (solid red line). The correlation time
for the red noise in the fit that is displayed is 3000 sec. The fit
improves slightly at low frequencies for longer correlation times,
but the data sampling is insufficient to constrain the red part of
the spectrum.}
\end{center}
\end{figure}
\section{Discussion} It was shown that two HPS decayed with e-folding
times of $T$ = 3.8 hrs and $T$ = 8.6 hrs at 15 GHz. This time scale
can be converted to a bulk velocity by considering the 15 GHz VLBA
images of three HPS. From these images, after the effective
beam-width was de-convolved, the empirically determined source of
the effective emissivity for the 15 GHz flux was found to decrease
exponentially along the jet direction with an e-folding distance of
1.2 mas $<D<$ 1.5 mas. Assuming that the power at the base is
abruptly diminished significantly and the previously ejected plasma
continues down the jetted path, an apparent bulk velocity was
estimated as $V_{app}=D/T$. It was found that $0.16c < V_{app} <
0.46c$. This transforms to an intrinsic velocity $0.17 c < v <
0.43c$.

\par We can compare these results to estimates of the bulk jet
velocity that arise from Doppler boosting and bilateral symmetry. In
\citet{dha00}, it was determined from these arguments that $v
\approx 0.1c$. In \citet{rib04}, using the same assumptions, $0.27c
< v < 0.54c$ based on later compact jet observations with VLBA.
These previous treatments assumed a larger angle of inclination,
$\theta =70^{\circ}$. However, $\theta = 60^{\circ}$ has now been
estimated from proper motion measurements of approaching and
receding ejecta during an outburst, combined with the
newly-determined parallax distance of 8.6 kpc \citep{rei14}. So
these older estimates are on the high side by a factor of
$\cos{60}/\cos{70}=1.47$. Thus, with the revised distance estimate
$0.18 c < v < 0.37c$ based on bilateral symmetry or $0.07 c < v <
0.37c$ if one also includes the \citet{dha00} results. There is an
important assumption in deriving $v$ in \citet{rib04}. The central
Clean Component (CC) was always assigned to the approaching jet.
This is roughly consistent with the location of the BHXRB being
upstream of the peak of the flux density. It was shown in Section 2,
based on our exponential line source model fits, that the peak flux
density lies 0.3 mas downstream of the beginning of the approaching
jet. This is roughly consistent with assigning the central CC to the
approaching jet. However, for completeness we note that if the
central CC is assigned to both jets or to neither of them, $v$
decreases to approximately 0.2c using $\theta =70^{\circ}$
\footnote{as communicated by M. Ribo at the EVN Symposium 2004 The
7th European VLBI Network Symposium on New Developments in VLBI
Science and Technology Toledo, Spain, 2004 October 12-15}. Thus, the
independent determination of $v$ in this paper and the bilateral
symmetry arguments provide excellent agreement. It might vary from
jet episode to jet episode, but in no case is it relativistic like a
radio loud quasar jet or even like the discrete (superluminal)
ejections.
\par The bilateral symmetry argument is more robust when taken as a
collected set of data as opposed to each observation individually.
If the intrinsic velocity were much larger than indicated by the
asymmetry in the images, it is hard to understand how different
epochs with different u-v coverage and observation durations could
consistently produce highly symmetric images. Namely, with many
observations, there is no reason to believe that there is somehow
always the right amount of missing flux due to poor u-v coverage
that skews the bilateral asymmetry to make it appear symmetric. This
argument is compelling whether the intrinsic source is actually
symmetric or not. Furthermore, our independent method (using
exponential decay lengths and times) of estimating $v$ from the VLBA
images implies $0.17c < v < 0.43c$, an almost identical range to
that implied by estimates from bilateral symmetry based on the same
radio images (using the new parallax determined line of sight) $0.18
c < v < 0.37c$ \citep{rib04}. It is concluded that the original
bilateral symmetry based estimates of $v$ from \citet{rib04} are
correct or there are numerous conspiring coincidences. The direct
implementation of VLBA images to constrain $v$ is in our opinion
superior to folding in information such as time scales of
quasi-periodic events (no evidence of quasi-periodicity was found in
Section 4) in order to estimate $v$.
\par It is natural to tie the X-ray observations of an HPS during decay in Section 3
to disk models of jet emission in BHXRBs. We concluded that during
strong jet emission there was a paucity of low energy Comptonizing
electrons. We also noted that the spectral power of X-ray variations
was at much lower frequency when the jet is at full force. If one
equates lower frequency to large dimensions this suggest a scenario
in which the low energy Comptonizing electrons occupy the inner
regions of the accretion flow plus corona. When the jet is fully
powered up it displaces these electrons. When the jet powers down
this region re-fills with low energy Comptonizing electrons. This
idea is similar to the unified model of GBH jets of \citep{fen04}.
The one difference is that instead of the jet displacing the
optically thick plasma responsible for the soft thermal emission
from the inner disk, the jet is displacing low energy optically thin
plasma from the inner regions of the accretion system. We do note
that if the disk temperature is too cold for us to identify it in
our spectral fits, a thermal disk that approaches the black hole
during the decay of the compact jet could also produce these
effects.

\par As promising as a unified picture of GRS~1915+105 jets appeared
to be as a laboratory for studying the more distant jets in radio
loud quasars (which evolve over time scales much longer than a human
lifetime), the popularized notion that Galactic black holes produce
highly relativistic jets was tempered by the first parallax distance
estimate to GRS~1915+105. Radio-loud quasar jets clearly propagate
with highly relativistic bulk velocities. However, the relativistic
speed inferred for discrete ejections in GRS~1915+105 was based on
crude distance estimates and physical arguments that yielded a
distance of 11 kpc - 12.5 kpc. The parallax determined distance
measurement is $8.6^{+2.0}_{-1.6}$ kpc \citep{rei14}. This changes
the intrinsic velocity of the discrete ejections to 0.65c - 0.81c.
As a corollary to this, it was shown that the analogy between the
discrete ejections and Fanaroff-Riley II RLQs is untenable if the
distance is less than 10.5 kpc \citep{pun13}. We note that the other
strong case for direct observational evidence of superluminal motion
in a Galactic black hole (GBH), GRO~J1655-40, was also based on a
crude distance estimate that has been questioned to be much smaller
based on newer data and these discrete ejections do not seem to be
superluminal \citep{fol06}. Based on the dichotomy of the types of
jetted emission noted in the Introduction, the lack of direct
observational evidence of highly relativistic motion in BHXRBs and
the lack of an analog with FRII jets indicates that the assumption
of relativistic bulk motion in compact jets should be used with
great caution. The findings of this study do not support highly
relativistic motion in compact jets as well. The analogy between
jets in BHXRBs and jets in radio loud active galactic nuclei is not
clear.

\end{document}